\documentclass[manuscript]{acmart}
\usepackage{hyperref}
\hypersetup{
 unicode=true,
 pdfstartview={FitH},
 colorlinks=true,
 citecolor=blue,
 linkcolor=blue,
 urlcolor=blue
}

\usepackage[linesnumbered,algoruled]{algorithm2e}
\usepackage{algpseudocode}

\usepackage{caption}
\usepackage{subcaption}
\usepackage{multirow}
\usepackage{makecell}
\usepackage[inline]{enumitem}
\usepackage{xcolor}
\definecolor{darkgreen}{RGB}{2,112,10}
\definecolor{darkpurple}{RGB}{131, 50, 168}
\usepackage{listings}
\lstdefinestyle{human}{
    basicstyle=\footnotesize\ttfamily\color{darkgreen},
    keywordstyle=\color{darkgreen},
    stringstyle=\color{darkgreen},
    commentstyle=\color{darkgreen},
    morecomment=[l][\color{darkgreen}]{\%},
    identifierstyle=\color{darkgreen},
    rulecolor=\color{black},
    numberstyle=\color{black}\ttfamily
}

\lstdefinestyle{llm}{
    basicstyle=\footnotesize\ttfamily\color{darkpurple},
    keywordstyle=\color{darkpurple},
    stringstyle=\color{darkpurple},
    commentstyle=\color{darkpurple},
    morecomment=[l][\color{darkpurple}]{\%},
    identifierstyle=\color{darkpurple},
    rulecolor=\color{black},
    numberstyle=\color{black}\ttfamily
}
\usepackage[comments]{annotations}

\newcommand{\nop}[1]{}

\newcommand{\Com}[1]{}

\crefname{algocf}{alg.}{algs.}
\Crefname{algocf}{Algorithm}{Algorithms}
\SetKwRepeat{Do}{do}{while}

\usepackage{enumitem}
\usepackage{acronym}
\acrodef{IP}[IP]{intellectual property block}
\acrodef{SoC}[SoC]{System-on-Chip}
\acrodef{IC}[IC]{integrated circuit}
\acrodef{eFPGA}[eFPGA]{embedded field programmable gate array}
\acrodef{FPGA}[FPGA]{field programmable gate array}
\acrodef{RTL}[RTL]{register transfer level}
\acrodef{CLB}[CLB]{configurable logic block}
\acrodef{LUT}[LUT]{look-up table}
\acrodef{HLS}[HLS]{high-level synthesis}
\acrodef{EDA}[EDA]{electronic design automation}
\acrodef{FF}[FF]{flip-flop}
\acrodef{DIP}[DIP]{distinguishing input pattern}
\acrodef{PPA}[PPA]{power, performance, and area}
\acrodef{BLE}[BLE]{basic logic element}
\acrodef{CB}[CB]{connection block}
\acrodef{SB}[SB]{switch block}
\usepackage{booktabs}
\newcolumntype{L}[1]{>{\raggedright\let\newline\\\arraybackslash\hspace{0pt}}m{#1}}
\newcolumntype{C}[1]{>{\centering\let\newline\\\arraybackslash\hspace{0pt}}m{#1}}
\newcolumntype{R}[1]{>{\raggedleft\let\newline\\\arraybackslash\hspace{0pt}}m{#1}}

\newcommand*\circled[1]{\tikz[baseline=(char.base)]{
            \node[shape=circle,fill,inner sep=1pt] (char) {\textcolor{white}{#1}};}}

\usepackage{amsmath,amsfonts}
\usepackage{url}

\AtBeginDocument{%
  }

\setcopyright{acmlicensed}
\acmJournal{TODAES}
\acmYear{2025} \acmVolume{1} \acmNumber{1} \acmArticle{1} \acmMonth{1}\acmDOI{10.1145/3734524}

\begin{document}

\title{C2HLSC: Leveraging Large Language Models to Bridge the Software-to-Hardware Design Gap}

\author{Luca Collini}
\email{lc4976@nyu.edu}
\author{Siddharth Garg}
\email{siddharth.garg@nyu.edu}
\author{Ramesh Karri}
\email{rkarri@nyu.edu}
\affiliation{%
  \institution{NYU Tandon School of Engineering}
  \city{Brooklyn}
  \state{New York 11201}
  \country{USA}
}

\renewcommand{\shortauthors}{Collini et al.}

\begin{abstract}
High-Level Synthesis (HLS) tools offer rapid hardware design from C code, but their compatibility is limited by code constructs. This paper investigates Large Language Models (LLMs) for automatically refactoring C code into HLS-compatible formats. We present a case study using an LLM to rewrite C code for NIST 800-22 randomness tests, a QuickSort algorithm, and AES-128 into HLS-synthesizable C. The LLM iteratively transforms the C code guided by the system prompt and tool's feedback, implementing functions like streaming data and hardware-specific signals. With the hindsight obtained from the case study, we implement a fully automated framework to refactor C code into HLS-compatible formats using LLMs. To tackle complex designs, we implement a preprocessing step that breaks down the hierarchy in order to approach the problem in a divide-and-conquer bottom-up way. We validated our framework on three ciphers, one hash function, five NIST 800-22 randomness tests, and a QuickSort algorithm. Our results show a high success rate on benchmarks that are orders of magnitude more complex than what has been achieved generating Verilog with LLMs.
\end{abstract}



\received[accepted]{30 April 2025}

\maketitle

\section{Introduction}

The increased demand for custom hardware accelerators shines a light on High-Level Synthesis (HLS) tools, which allow the fast design of accelerators by converting C code into hardware description languages (HDLs) such as Verilog and VHDL. HLS tools convert a high-level specification (C, C++) into an \ac{RTL} description~\cite{hls}: (1) HLS uses state-of-the-art compilers (e.g., LLVM or GCC) to extract a high-level control data flow graph (CDFG). (2) They then assign operations to time (scheduling) and space (allocation and binding) to determine the micro-architecture. HLS tools support pragmas and directives to explore architectural choices for a C specification. HLS, though, does not come without its shortcomings. In fact, it can work on a subset of C code and often requires reformatting code in specific ways that are easier to map into hardware. This is because software and hardware paradigms are different. 
For instance, hardware does not support dynamic memory allocation and recursive constructs. Outputs can only communicate through parameters, array sizes need to be static, limited support for pointers, and multiple processes can be modeled through independent function instances mapped into hardware blocks. Designers manually refactor C code to remove these constructs and make it compatible with HLS tools. Such manual refactoring is time-consuming and error-prone~\cite {deeprajQPC}. We explore the use of Large Language Models (LLMs) to aid developers in porting generic C code into HLS synthesizable C code. LLMs can write Verilog code~\cite{chipchat}. Since the available Verilog code is very scarce, compared to software languages like C and Python, LLMs perform poorly on HDLs. By raising the level of abstraction, we propose to use LLMs to write HLS C and aid hardware design effectively.  We build upon the work  in~\cite{c2hlsc}, expanding the prototype flow into a complete flow. The main contributions are:
\begin{itemize}
    \item An automated flow for rewriting C code into synthesizable C code, which:
    \begin{itemize}
        \item Supports hierarchical designs, automatically building unit tests from a top-level test;
        \item Supports function rewriting for streaming functions;
        \item Supports pragma generation to target either area or throughput objectives;
        \item Uses in-context learning and smart model selection to reduce cost;
    \end{itemize}
    \item An experimental evaluation using real-world applications benchmarks, including crypto and NIST implementations of TRNG tests.
\end{itemize}

The paper road-map is as follows: 
\begin{enumerate}
    \item \autoref{sec:rel} gives an overview of High-Level Synthesis tools and large language models.
    \item \autoref{sec:rel_works} summarizes related works, highlighting the novelties of our approach.
    \item \autoref{sec:cs} illustrates the case study of an engineer-in-the-loop approach, presenting the tasks, methodologies, and results motivating the development of a fully automated framework.
    \item \autoref{sec:automated} presents the fully automated C2HLSC tool, discussing capabilities and limitations.
    \item \autoref{sec:exp_eval} reports the experimental setup and analyzes and discusses the results.
    \item \autoref{sec:conc} concludes the paper . 
\end{enumerate}

\section{Background} \label{sec:rel}
\subsection{High-Level Synthesis: capabilities and limitations}
\label{sec:hls}
HLS  methods automatically convert a high-level software specification into a corresponding \ac{RTL} description~\cite{hls}. 
The resulting component is a specialized IP block tailored to execute the functionality. The HLS process is based on state-of-the-art compilers (e.g., LLVM or GCC), which extract a high-level representation of the functionality and assigns the corresponding operations to time (scheduling) and space (allocation and binding) to determine the micro-architecture. Engineers can instruct the HLS tool to adopt different implementation strategies for different parts of the code by using pragmas or directives. Pragmas and directives formats are tool-specific, but most tools support the following optimizations:
\begin{itemize}
    \item \textbf{Loop unrolling}: multiple iterations of a loop are executed simultaneously, reducing the number of loop iterations. In hardware, this increases parallelism by replicating hardware resources for each unrolled iteration. For example, instead of processing one element per clock cycle, an unrolled loop might handle multiple elements concurrently, improving throughput for additional area due to resource duplication.
    \item \textbf{Loop/Function pipelining}:
    overlaps the execution of consecutive loop iterations/function calls, much like instruction pipelining in processors. This allows new iterations to begin before the previous ones complete, improving latency and resource utilization. The pragma controls the initiation interval (II), which determines how often new loop iterations start. A smaller II leads to higher throughput but requires more resources.
    \item \textbf{Array partitioning}:
    is used to divide arrays into smaller, independent memory blocks, allowing parallel access. This alleviates memory access bottlenecks by enabling multiple elements to be read from or written to simultaneously. Different partitioning schemes, such as block or cyclic partitioning, can be applied depending on the access patterns in the code. This improves bandwidth and efficient memory usage. 
\end{itemize}
Although other pragmas exist for defining module interfaces and other architectural optimization, these are the most commonly used pragmas and the ones that we focus on in the optimization part of this framework.

HLS has its shortcomings. Software paradigms assume the code to be executed on a Von Neumann machine, capable of dynamic memory allocations, stack management for function calls, and often with an operating system that provides routines for executing multiple processes. All these language features cannot be mapped into a hardware digital circuit and, for this reason, are not supported by HLS tools. 
The following are the most common HLS limitations:
\begin{enumerate}
    \item \textbf{Dynamic Memory Allocation}:
    \begin{itemize}
        \item \textbf{Description}: Functions like \texttt{malloc}, \texttt{calloc}, \texttt{realloc}, and \texttt{free} are used for dynamic memory allocation in C. Hardware requires fixed memory allocation, making these functions incompatible.
        \item \textbf{Example}:
        \begin{lstlisting}[language=c]
        int *arr;
        arr = (int *)malloc(x * sizeof(int));
        if (arr == NULL) {
            // Handle memory allocation failure
        }
        \end{lstlisting}
        This C code allocates memory X integers and assigns the address of the allocated memory to arr.
        To make this code compatible with HLS tools, the engineer should identify an upper bound for the number of integers needed and use a fixed-size array.
    \end{itemize}
    \item \textbf{Recursion}:
    \begin{itemize}
        \item \textbf{Description}: Recursive functions require dynamic stack management, which is not present in hardware. HLS tools build a hardware module for each function. A module that contains itself is not feasible.
        \item \textbf{Example}:
        \begin{lstlisting}[language=c]
        int factorial(int n) {
            if (n == 0) return 1;
            else return n * factorial(n - 1);
        }
        \end{lstlisting}
        Every recursive function can be rewritten in an iterative version. For the general case, it is necessary to implement a stack, and the engineer should identify an upper bound for it.
    \end{itemize}
    \item \textbf{Pointers}: Pointers, especially those involving dynamic memory or complex pointer arithmetic, have limited support in hardware as there is no concept of addressable memory.    
    \item \textbf{Standard Library Functions} Many standard C library functions, especially those related to I/O, file handling, and dynamic memory, are not supported.        
    \item \textbf{Complex Data Structures}:
    dynamic or nested elements can be challenging to map directly to hardware.      
    \item \textbf{Multiple Processes}:
    Multiple processes or threads are not directly supported; instead, independent function instances are used to model parallelism.        
\end{enumerate}

\subsection{Large Language Models (LLMs)} 
LLMs are trained on vast amounts of text data, excelling in tasks such as code generation and translation, particularly in widely used programming languages like C, C++, and Python. However, they encounter challenges when applied to HDLs like Verilog or VHDL, due to the comparatively limited amount of training data in these specialized languages~\cite{hammond2022}. In light of this, the focus of this paper shifts from generating HDL code to leveraging LLMs for refactoring C code into a subset that is compatible with HLS tools.
Rather than targeting Verilog or VHDL directly, this approach capitalizes on HLS, which enables the design of hardware systems from C code. However, HLS imposes strict requirements, only functioning with specific code constructs as in \autoref{sec:hls}. LLMs, with their proven strengths in understanding and transforming general-purpose C code, can be used to refactor this code into an HLS-synthesizable form while maintaining its original functionality. We provide instructions and examples in the form of in-context learning (ICL) to instruct the LLM on the kind of transformations needed for HLS. This strategy allows for a streamlined hardware design process by using LLMs' capabilities in code manipulation, bypassing their limitations in HDL generation.

\section{Related works}\label{sec:rel_works}
Previous work explored LLMs to design hardware. Verigen fine-tuned an LLM to improve its ability to produce Verilog~\cite{verigen}. The fine-tuned LLM, though, performed marginally better than ChatGPT3.5-turbo with an accuracy $\sim$65\%.  ChipChat~\cite{chipchat} was the first to tapeout a design written by an AI model. However, the single-shot performance of the AI model was low and needed several iterations in order for the LLM to get to the correct result. In AutoChip~\cite{autochip}, authors proposed the use of simulation-based feedback to automatically guide the LLM towards a correct design iteratively. The approach resulted in success, but only on very simple benchmarks. By leveraging C code generation, we aim to validate our work on real-world applications. In~\cite{chatchisel} Liu et al. showed that raising the level of abstraction to Chisel (a domain-specific language based on Scala) improved the LLM success rate by 30\% with respect to Verilog generation. 
We target generating synthesizable C code as LLMs are more capable at C than at hardware languages~\cite{hammond2022}. In~\cite{amaranth}, an LLM was used to write Amaranth HDL, a Python-based HDL that allows the modeling of synchronous logic at the RTL. For this reason, while it uses a high-level language, its semantics are close to Verilog and is targeted to hardware designers. While the LLM came up with parts of the design, it fell short in some tasks, like generating interfaces. 
Software developers use HLS to design hardware, and as such, the code only provides the functionality\footnote{Whereas the hardware architecture and interface specification are instructed using HLS pragmas and directives.}. The first LLM-based approach to generating accelerators leveraging HLS was GPT4AIGChip\cite{GPT4AIGChip}. The GPT4AIGChip framework is based on user-provided templates to generate C code and directives to build accelerators. The process is not fully automated as some limitations in the framework (interfaces and composing different functions) need to be overcome by hand. Liao et al.~\cite{llm_hls1} evaluated natural language to synthesizable C and C to Verilog using LLMs. Similarly, Swaroopa et al.~\cite{eval_llm_hls} explored natural language to synthesizable C generation, finding frequent problems in the correctness of the generated code. 
We focus on transforming generic C to synthesizable C using LLMs. Our gap is closer, and our inputs and outputs are of the same nature, making this approach more promising to succeed. At the same time, our approach is relevant because most new algorithms are first implemented in software (i.e., reference implementations of NIST standards) and then need to be accelerated in hardware. Moreover, we can leverage the existing testing infrastructure to verify the correctness of the LLM-generated code.  Our approach leverages the existing software code bases to build their hardware accelerators. HLSPilot~\cite{llm_hls2} is a framework that focuses on kernel identification and optimization starting from C/C++ code. To the best of our knowledge, HLSPilot does not focus on code repairs for synthesizable code. 
In~\cite{automatedhls} Xu et al. proposed a framework to automatically repair C code for HLS. They first use state-of-the-art techniques to repair simpler issues and then use an LLM to repair the remaining issues. Their results show that their approach struggles on bigger benchmarks (for example, the success rate on AES is 60\%, compared to 80\% in our work). We include a preprocessing step to break down the design hierarchy and tackle the design in a bottom-up divide-and-conquer fashion, tasking the LLM with more, but simpler tasks. 

\section{Interactive Approach: a Case Study}\label{sec:cs}
\subsection{Overview}
\begin{figure}[h]
  \centering
  \includegraphics[width=0.8\columnwidth]{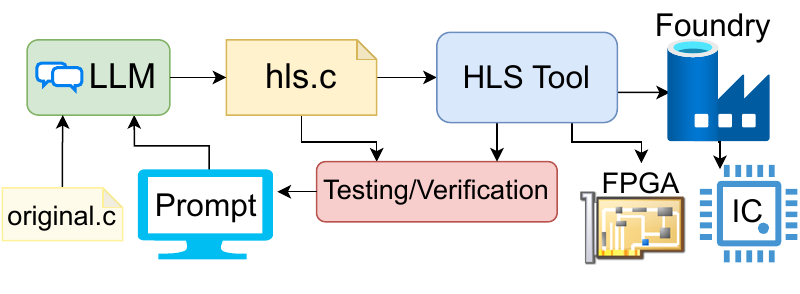}
  \vspace*{-15pt}
  \caption{Flow used in the C2HLSC case-study.}
  \vspace*{-15pt}
  \label{fig:hlsflow}  
  
\end{figure}

We performed a case study manually prompting Gemini LLM~\cite{gemini} to refactor C code into synthesizable C suitable for HLS. The goal of this case study was to explore the potential and limitations of LLM in refactoring C code for HLS tools. The evaluation consisted of two tasks. The {\bf first task} involved rewriting reference C implementations of the Frequency test, Frequency Block test, Cumulative Sums, and Overlapping Template Matching tests from the NIST 800-22 suite~\cite{nist} into synthesizable C code. These tests are designed to assess the randomness of a sequence. A {\bf first challenge} arose due to the inherent differences between software and hardware implementations. The reference C implementations operate on a pre-loaded random sequence stored in memory. Conversely, hardware implementations require on-the-fly analysis, processing the sequence bit-by-bit. This necessitates modifying the code to handle a streaming data input rather than a pre-loaded array. A {\bf second challenge} stemmed from the p-value calculation. In the software context, the precise p-value is critical and computed on the fly. However, since the hardware implementations primarily focus on distinguishing random from non-random sequences, this process can be simplified by pre-computing certain values offline and reducing the computational burden during on-the-fly analysis. Both these challenges – adapting to streaming data and simplifying p-value calculations – are non-trivial for human developers and LLMs. 
The {\bf second task} assesses the LLM's ability to rewrite code constructs that are not supported by HLS tools. We used two algorithms: a QuickSort containing pointers and recursion~\cite{quicksort}, and the AES128 encrypt from the tinyAES library~\cite{tinyAES} with six functions. The goal was for the LLM to generate code without pointers and recursion, making it suitable for (Catapult) HLS.

\subsection{Methodology}
\autoref{fig:hlsflow} illustrates our 
We broke down the process into small steps to allow the LLM to transform the original C into synthesizable C. For the first task, we followed the following steps for the three tests: 
\begin{enumerate*}
    \item Present task to the LLM: "Hi, I have this code in C that I need to rewrite such that I can use it with an HLS tool to generate hardware.".
    \item Ask to remove print statements.
    \item Ask to rewrite the function as a streaming interface: "Now I need to rewrite the function such that it will get inferred as a streaming interface, to do so, I need to get rid of the epsilon array and have the function take a parameter to accept a bit at each function call."
    \item Ask to remove math steps to be computed offline (in some cases, ask to write a script to run them).
    \item Ask to add \textit{is\_random} and \textit{valid} signals as parameters.
    \item Ask to optimize data types using arbitrary width integers and fixed point arithmetic using HLSLIBS~\cite{hlslibs}.
    \item Ask to write a main function to test the function passing random bits. 
    \item Ask to fix mistakes passing errors from the HLS tool.
\end{enumerate*}

For QuickSort, we followed these steps:
\begin{enumerate*}
    \item Present the task to the LLM: "Hi, I have this code in C that I need to rewrite such that I can use it with an HLS tool to generate hardware.".
    \item Ask to remove print statements.
    \item Ask to rewrite function without using pointers.
    \item Ask to rewrite function without recursion.
    \item Ask to fix array sizes in function parameters.
    \item Ask to optimize data types using arbitrary width integers and fixed point arithmetic using HLSLIBS.
    \item Ask to write a main function to test the function passing an array to sort.
    \item Ask to fix mistakes by passing errors from the HLS tool.
\end{enumerate*}

For the AES 128 from tinyAES~\cite{tinyAES} we followed the following steps asking to fix one function at a time:
\begin{enumerate*}
    \item Present the task to the LLM: "Hi, I have this code in C that I need to rewrite such that I can use it with an HLS tool to generate hardware.".
    \item Ask to rewrite for loops with fixed bounds and no pointer usage.
    \item Ask to rewrite the function parameters to use fixed-size arrays.
    \item Ask to fix eventual mistakes passing errors from the HLS tool.
\end{enumerate*}
When the LLM responds with sub-optimal answers, we check alternative answers. If none fully satisfy the request, we instruct the LLM with additional prompts, including more details pointing out where the problem was and, if not sufficient, hinting at possible solutions.

\subsection{Case Study Results} \label{sec:res}
This study aimed to evaluate how LLMs perform at rewriting C code so that it is HLS synthesizable. We run the code through Catapult HLS to check correctness after synthesis, but we do not focus on resource utilization, as it depends on the architectural decisions. We targeted the nangate45 library at 50 MHz with a synchronous active high reset for all the tests.
The LLM was able to rewrite all C code to run on Catapult HLS. We performed simulations with Modelsim to check the equivalency of the result between the original C and the synthesized Verilog obtained from the LLM-generated C. We can classify the errors in the LLM-generated code into compile/functional errors and synthesis errors. The syntax and synthesis errors were easier to fix, instructing the LLM with the error message and the affected line. For functional errors caught by the failing tests, it was harder for the LLM to identify the cause and propose a fix. Manual effort was needed to point out the sources of the issues so that the LLM could propose an effective fix. Typical functional issues were mistaken loop readjustments, widths, and arbitrary width types.

\subsubsection{C2HLSC of NIST 800-22 Randomness Tests}
The LLM performed well in most steps, especially in rewriting the function as a streaming interface, which is one of the more complex steps. Figure \ref{fig:streaming} shows the streaming interface rewriting for the Frequency Block test. 
For all three tests, this step was performed correctly at the first shot. However, the LLM did stumble on tasks that seemed more trivial. For example, when asked to remove the \texttt{abs} function in this line of code \texttt{(abs(sum.to\_int())  \textless= SUM\_THRESHOLD);} it came up with convoluted ways that did not work like \texttt{(sum.to\_int() \textless 0) ? -sum.to\_int() : sum.to\_int()) \textless= SUM\_THRESHOLD} and it took many tries to eventually get the expected result \texttt{(sum \textless SUM\_THRESHOLD) \&\& (sum \textless -SUM\_THRESHOLD)}. 
The test function generation was always generated effectively. 
\begin{figure}[thb]
  \centering
  \begin{tabular}{cc}
  \begin{subfigure}[t]{.44\columnwidth}
    \begin{lstlisting}[style=human]
I need to rewrite the function such that it 
will get inferred as a streaming interface, 
to do so, I need to get rid of the epsilon 
array and have the function take a parameter
to accept a bit at each function call.    
\end{lstlisting}
\vspace{-7pt}
  \end{subfigure}
  & \hspace{4pt}
  \begin{subfigure}[t]{.44\columnwidth}
    \begin{lstlisting}[style=llm]
Here's the code rewritten to accept a single
bit per function call, making it more 
suitable for a streaming interface:
    \end{lstlisting}
    \vspace{-7pt}

  \end{subfigure} \\

  \begin{subfigure}[t]{.44\columnwidth}
    \begin{lstlisting}[language=c]
void BlockFrequency(int M, int n) {
int i, j, N, blockSum;
double p_value, sum, pi, v, chi2;
N = n / M;  /* # OF SUBSTRING BLOCKS */
sum = 0.0;
for (i = 0; i < N; i++) {
    blockSum = 0;
    for (j = 0; j < M; j++) {
        blockSum += epsilon[j + i * M];
    }
    pi = (double)blockSum / (double)M;
    v = pi - 0.5;
    sum += v * v;
}
chi2 = 4.0 * M * sum;
p_value = cephes_igamc(N / 2.0, chi2 / 2.0);}
\end{lstlisting}
\vspace{-7pt}
  \end{subfigure}& \hspace{4pt}
  \begin{subfigure}[t]{.44\columnwidth}
    \begin{lstlisting}[language=c]
void BlockFrequency(int M, int n, bool bit) {
static int i = 0, j = 0, blockSum = 0;
static double sum = 0.0;
double pi, v, chi2;
if (j == 0) { // Reset on new block
    i++;
    blockSum = 0;
}
blockSum += bit; // Accumulate bit
j++;
if (j == M) { // Block complete
    pi = (double)blockSum / (double)M;
    v = pi - 0.5;
    sum += v * v;
    j = 0;
}
if (i == (n / M)) { // All blocks processed
    chi2 = 4.0 * M * sum;
    p_value=cephes_igamc(n/2.0, chi2/2.0);
}}
\end{lstlisting}
\vspace{-7pt}
  \end{subfigure}
  \end{tabular}
  \caption{Transforming into a streaming pipeline. On the left, in green, human prompt. On the right, in purple, LLM response}
  \label{fig:streaming}
\end{figure}

\subsubsection{C2HLSC of QuickSort}
The original QuickSort used a function to swap elements in the arrays using pointers. Inlining the function and removing pointers was flawless, as reported in Figure \ref{fig:pointers}. However, in the subsequent steps it reverted to using the function and needed re-prompting to inline.
The LLM eliminated recursion in one shot, using a fixed-size stack to track the high and low values. However, the stack management logic was not correct. The problem was identified at the end when running the generated test, which caused a segmentation fault. The LLM fixed the problem in a few tries by feeding the errors from gdb and showing intermediate values of high and low.

\begin{figure}[thb]
  \centering
  \begin{tabular}{cc}
  \begin{subfigure}[t]{.44\columnwidth}
    \begin{lstlisting}[style=human]
Can we get rid of the swap function and 
replace it inline? 
\end{lstlisting}
\vspace{-7pt}
  \end{subfigure} & \hspace{4pt}
   \begin{subfigure}[t]{.44\columnwidth}
    \begin{lstlisting}[style=llm]
Yes, you can definitely get rid of the swap
function and replace it inline within the
partition function. Here's the modified code:
    \end{lstlisting}
    \vspace{-7pt}

  \end{subfigure}\\
  \begin{subfigure}[t]{.44\columnwidth}
    \begin{lstlisting}[language=c]
void swap(int* a, int* b) {
    int temp = *a;
    *a = *b;
    *b = temp;}
...
    if (i < j) {
        swap(arr[i], arr[j]); 
    }}
swap(&arr[low], &arr[j]);
...
\end{lstlisting}
\vspace{-7pt}
  \end{subfigure}
  & \hspace{4pt}
  \begin{subfigure}[t]{.44\columnwidth}
    \begin{lstlisting}[language=c]
...
    if (i < j) {
        // Inline swap using temporary variable
        int temp = arr[i];
        arr[i] = arr[j];
        arr[j] = temp;
    }}
// Swap pivot element with element at index j
int temp = arr[low];
arr[low] = arr[j];
arr[j] = temp;
return j;
...
\end{lstlisting}
\vspace{-7pt}
  \end{subfigure}
  \end{tabular}
  \caption{Removing pointers and inlining. On the left, in green, human prompt. On the right, in purple, LLM response.}
  \label{fig:pointers}
\end{figure}

\subsubsection{C2HLSC of AES 128 Encrypt}
We fed one function at a time to the LLM starting from the top function \texttt{Cipher}. We tested the generated C code and found errors. We prompted the LLM to correct the errors. The LLM made major changes in the MixColumn and ShiftRows functions, so we asked it to double-check that the new code was correct. The ShiftRow function was corrected in 2 iterations without extra context on the error. Results are in Figure \ref{fig:aes}. Fixing the MixColumn function took more iterations even though the error was simple.
\begin{figure}[tb]
  \centering
  \begin{tabular}{cc}

  \begin{subfigure}[t]{.44\columnwidth}
    \begin{lstlisting}[style=human]
Can you fix possible problems in 
this function?
\end{lstlisting}
\vspace{-7pt}
  \end{subfigure}
  & \hspace{4pt}
  \begin{subfigure}[t]{.44\columnwidth}
    \begin{lstlisting}[style=llm]
LLM result after 2 tries telling the LLM that 
the loop logic was wrong.
    \end{lstlisting}
    \vspace{-7pt}
  \end{subfigure}
  \\
  \begin{subfigure}[t]{.44\columnwidth}
    \begin{lstlisting}[language=c]
static void ShiftRows(state_t* state){
  uint8_t temp;
  // Rotate first row 1 columns to left  
  temp           = (*state)[0][1];
  (*state)[0][1] = (*state)[1][1];
  (*state)[1][1] = (*state)[2][1];
  (*state)[2][1] = (*state)[3][1];
  (*state)[3][1] = temp;
  // Rotate second row 2 columns to left  
  temp           = (*state)[0][2];
  (*state)[0][2] = (*state)[2][2];
  (*state)[2][2] = temp;
  temp           = (*state)[1][2];
  (*state)[1][2] = (*state)[3][2];
  (*state)[3][2] = temp;
  // Rotate third row 3 columns to left
  temp           = (*state)[0][3];
  (*state)[0][3] = (*state)[3][3];
  (*state)[3][3] = (*state)[2][3];
  (*state)[2][3] = (*state)[1][3];
  (*state)[1][3] = temp;}
...
\end{lstlisting}
\vspace{-7pt}
  \end{subfigure}
  & \hspace{4pt}
  \begin{subfigure}[t]{.44\columnwidth}
    \begin{lstlisting}[language=c]
static void ShiftRows(state_t state[4][4]) {
uint8_t temp;
int i, j;
// Loop over rows 1 to 3 (0-based indexing)
for (i = 1; i < 4; ++i) {
// Number of positions to left shift for current row
  int shift = i;
  // Use a separate loop for the number of shifts
  for (j = 0; j < shift; ++j) {
    temp = state[0][i];  // Store the element to be shifted
    // Shift all elements by one position to the left
    for (int k = 0; k < 3; ++k) {
      state[k][i] = state[k + 1][i];
    }
  state[3][i] = temp; // Place the temp value at the end
}}}
\end{lstlisting}
\vspace{-7pt}
  \end{subfigure}
    \end{tabular}
  \caption{Removing pointers. On the left, in green, human prompt. On the right, in purple, LLM response}
  \label{fig:aes}
\end{figure}

\subsection{Case Study Discussion}
Using LLMs to refactor C code into HLS-compatible formats is a promising avenue in LLM-aided design. The LLM, in our case Google's Gemini, refactored reference C implementations of NIST 800-22 randomness tests, QuickSort, and AES-128. The LLM's ability to iteratively refactor the code based on user instructions significantly reduces manual effort in adapting code for HLS tools. This can streamline the hardware design process, especially for repetitive tasks. The LLM effectively addressed challenges like converting code from memory-based data processing to streaming, from recursion to iteration and pointers. While the LLM achieved core functionalities, it occasionally struggled with minor details, requiring several iterations to guide it to the correct solution. In a practical scenario, a developer can rectify these minor errors. However, for an automated flow, a feedback loop is crucial, like that in~\cite{autochip}.

\begin{table}[!bht]
\centering
\caption{Resource Utilization and Latency Results. Typical operations include memory read and writes and basic mathematical operations (such as XOR, sum, subtraction). For Monobit the operations are memory reads/writes, sums and comparators. The different numbers of operations between manual and LLM Assisted depend on unrolling/pipelining directives.}
\label{tab:area}
\begin{tabular}{@{}lrrrrrr@{}}
\toprule
\multirow{2}{*}{Design} & \multicolumn{2}{c}{Area Score}                             & \multicolumn{2}{c}{\# Operations}                                  & \multicolumn{2}{c}{Latency}                                \\
                        & \multicolumn{1}{l}{LLM Assisted} & \multicolumn{1}{l}{Manual} & \multicolumn{1}{l}{LLM Assisted} & \multicolumn{1}{l}{Manual} & \multicolumn{1}{l}{LLM Assisted} & \multicolumn{1}{l}{Manual} \\ \midrule
NIST-Monobit  & 244  & 225.3  & 19 & 19  & 1 & 1  \\
NIST-Monobit Block  & 702.3   & 826.0  & 24  & 20  & 1  & 1  \\
NIST-Cusums  & 677.4  & 632  & 24 & 28 & 1   & 1  \\
NIST-Overlapping & 9933.4 & 7172.1  & 165 &  118 & 1 & 1 \\ 
QuickSort  & 18115.8 & n.a.  & 67 & n.a. & 18   & n.a. \\
AES & 38604.5  & n.a.  & 1924 & n.a. & 160  & n.a. \\ \bottomrule
\end{tabular}%
\end{table}

Table \ref{tab:area} shows the area for the implemented designs. For NIST test implementation, we have reference designs that were implemented by a graduate student. We used the same directives for a fair comparison between the 2. Area scores from Catapult are close. The manual implementations took around 4 hours each, while C2HLSC took between 30 to 60 minutes each. Although the sample size is limited, this shows the potential of LLMs to speed up the process effectively and efficiently.

\section{Fully Automated C2HLSC Framework}\label{sec:automated}
\begin{figure}[!htb]
  \centering
  \includegraphics[width=\columnwidth]{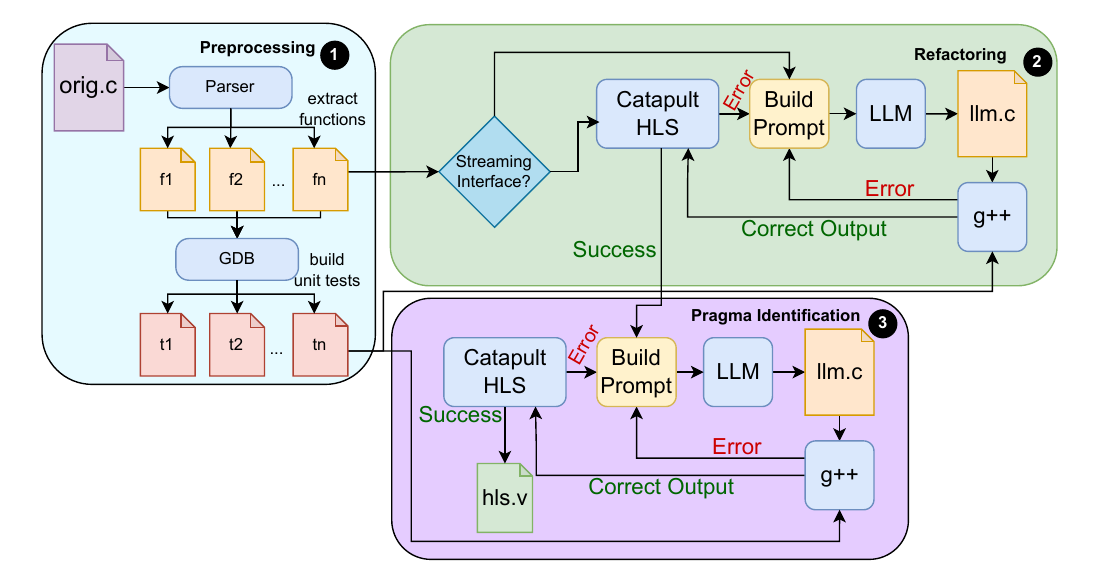}
  \caption{Automated C2HLSC LLM-based prototype flow.}
  \label{fig:prototype}
\end{figure}

With the experience of the engineer-in-the-loop case study, we implemented a fully automated C2HLSC prototype. An overview of the framework is provided in Figure~\ref{fig:prototype}.
The first insight from the case study was that the LLM can handle single-function tasks but has difficulties when prompted to work on multiple functions simultaneously. For this reason, we implemented a pre-processing step to handle hierarchical designs. A second insight was the twofold nature of the errors that can occur in the generated C: functional/compile errors and synthesis errors. The former can be caught by the inner loop (G++ and running unit tests). The latter can be caught by running the HLS tool. For this reason, we set up a double feedback loop as shown in Figure~\ref{fig:prototype}. One checks that the generated code compiles and passes reference tests, and one checks that the code is synthesizable by Catapult HLS.
Once the code is synthesizable, we enter a new step to apply pragmas to the code. This step is based on the same double-loop structure of the code refactoring step. We selected OpenAI and Anthropic models as we did not have access to Gemini APIs. Our framework is built around a model, the model itself can be easily swapped as more powerful models get released. The flow is implemented in Python and is available at \href{https://github.com/Lucaz97/c2hlsc}{C2HLSC}.

\subsection{Hierarchical Preprocessing}
From our engineer-in-the-loop case study, we learned that prompting the LLM to fix the whole code base at once was not effective for larger code bases. For this reason, we implemented a preprocessing step (illustrated in \autoref{fig:prototype} \circled{1}) that has the role of breaking down the design hierarchy so that the code refactoring step can be applied to one function at a time. We parse the input C code and identify the function hierarchy starting from the top function specified by the user. The code refactoring step requires a test to verify functional correctness. To ease the engineer, we only require a top-level test and automatically build unit tests for the child functions as part of the hierarchical preprocessing step. To identify the inputs to the child functions, we compile the code in debug mode and run it with gdb. We set a breakpoint for each function call and read out the parameters value. We then use these values to build unit tests for the child functions. In this way, we can test the inner functions with the same values they would be called with when testing the top-level function. The whole process is automated with a Python script.

\subsection{Code Refactoring}\label{sec:refactor}
The code refactoring step (illustrated in \autoref{fig:prototype} \circled{2}) is the heart of the C2HLSC flow. This step works on a single function. Functions are provided from the hierarchical preprocessing step, starting from leaf functions and then going up the hierarchy; in this way, we can always compile and synthesize the current function to test it. We noticed that LLMs have the tendency to get "distracted" by the function calls. Very often, the LLM would respond by reimplementing a function or providing a new signature for the function (a function signature consists of the function's name, its parameters and their data types, and its return type), and the provided code would not compile when integrated with the results from the child functions. Sometimes, this was also caused by the LLM changing the signature of the child functions, which would then not be compatible with the parent calls. For this reason, we provide in the prompt the signature of the child functions from the previous iterations and instruct the LLM to consider those as provided.
The code refactoring step may begin with an optional prompt to refactor the code to obtain a streaming interface. If this is not needed, the input function is synthesized with the HLS tool to identify the first error to fix. We build a prompt starting from the HLS tool error. Our system and initial prompts are listed in \autoref{fig:c2hlsc_prompt}. For the most common errors, we provide in-context learning examples to fix the error in the prompt. \autoref{sec:refactoring_icl} reports the in-context learning examples that we provide in our prompts.

\begin{figure}[h]
\centering
\begin{subfigure}{\columnwidth}
    \begin{lstlisting}
You are a C and High-Level Synthesis (HLS) expert. 
Assist with coding tasks to produce synthesizable HLS code. 
Your response should include a C code snippet that only modifies the specified functions, 
while leaving the rest of the code unchanged.  Do not add pragmas or directives, 
and ensure the code allows the HLS tool to infer the correct behavior.
\end{lstlisting}
\end{subfigure}
\begin{subfigure}{\columnwidth}
    \begin{lstlisting}
Help me rewrite the <top_function> function to be compatible with HLS: 
```
<code_to_fix>
``` 
The following child functions and includes will be provided with the following signature, assume them present in the code:
```
<includes>
<signatures>
```
The current problem is:" 
<error_from_catapult>
also include a main function that tests the code in the same way of the reference code:
```
<test_code>
```
\end{lstlisting}
\end{subfigure}
\vspace{-7pt}
  \caption{System prompt (on top) and initial prompt (on bottom) for code refactoring.}
  \label{fig:c2hlsc_prompt}
\end{figure}

We implemented a double feedback loop to guide the LLM to refactor the original C code into synthesizable C code. 
The inner loop uses g++ to compile the code. If the compilation process fails, we build a prompt including the error message from g++, and we prompt the LLM with it. This inner loop allows us to catch syntax and functional errors very quickly, as g++ is orders of magnitude faster than running the HLS tool directly. When the code compiles and runs successfully, we synthesize it with the HLS tool. If the synthesis succeeds, we proceed with the pragma identification step; otherwise, we build a new prompt as described above and re-iterate the loop.

\subsection{Pragma Identification}
The pragma identification step (illustrated in \autoref{fig:prototype} \circled{3}) has the same structure as the code refactoring step, but we changed the task for the LLM. In this step, we prompt the LLM to add pragmas to the code either to reduce the area or maximize throughput. We use in-context learning to provide the available pragmas and syntax of the HLS tool. We need the double feedback loop as sometimes the LLM will change the code even if the system prompt asks to only add pragmas. For this reason, we need to check that the code still behaves as expected before we synthesize it. As reflected by our system prompt listed in \autoref{fig:opt_prompt}, we ask the LLM to focus the optimizations around loop unrolling, pipelining, array partitioning, and function inlining. We found that using this strategy, the LLM is less prone to change the code, possibly entering loops of error fixing. This approach allows us to get a higher success rate while still using the most common optimization pragmas.

\begin{figure}[h]
\centering
\begin{subfigure}{\columnwidth}
    \begin{lstlisting}
You are a C and High-Level Synthesis (HLS) expert. Assist in coding tasks aimed at 
optimizing synthesizable HLS code. Your response must include a C code snippet that 
modifies only the specified functions for optimization. Do not change functionality,
and only add pragmas without modifying the function logic.
Optimize the code for either area or latency as instructed. 
Possible optimization mechanisms include:
    Loop Unrolling: Use "#pragma hls_unroll X" to unroll loops with a factor of X. 
                    Set X to yes to fully unroll the loop. Unrolling reduces latency
                    at the cost of area.
    Pipelining: Use "#pragma hls_pipeline_init_interval X" with X as the initiation 
                interval to pipeline loops. 0 disables pipelining. Pipelining 
                can be applied to loops to increase throughput at cost of latency.
If no optimization is needed, simply rewrite the original function.
\end{lstlisting}
\end{subfigure}
\begin{subfigure}{\columnwidth}
    \begin{lstlisting}
Update the <top_function> function to optimize it for HLS targeting <area|latency>.
The function is
```
<code_to_optimize>
```
The following child functions and includes will be provided with the following
signature, assume them present in the code:
```
<includes>
<signatures>
```
You should include a main function that tests the code in the same way of the
reference code: 
```
<test_code>
```
\end{lstlisting}
\end{subfigure}
\vspace{-7pt}
  \caption{System prompt (on top) and initial prompt (on bottom) for code optimization.}
  \label{fig:opt_prompt}
\end{figure}

\section{Experimental Evaluation}\label{sec:exp_eval}
We implemented the C2HLSC framework in Python, using pycparser\cite{pycparser} to parse the input C code and generate unit tests code in the hierarchical preprocessing step.  Our framework is available at \href{https://github.com/Lucaz97/c2hlsc}{C2HLSC}. We selected Catapult HLS as our high-level synthesis tool as it is an industry-level tool capable of targeting both ASIC and FPGA flows. We targeted nangate45 at 50MHz with synchronous active high reset for all runs. We selected models from OpemAI and Anthropic. In particular, we employ an ensemble approach using \textit{ChatGPT 4o-mini} and \textit{4o}~\cite{gpt4o}. We begin our flow with the smaller and cheaper 4o-mini model, and we switch to the more advanced but more expensive \textit{4o} model after 3 failing iterations. In this way, if the problem at hand is simple, we can solve it efficiently with the smaller model and use the more complex one only for the more challenging tasks. In this way, we save both execution time and cost as the smaller model is cheaper and faster than the more complex one.
We also use \textit{Claude Sonnet 3.5}~\cite{claude35} from Anthropic for a comparison. We evaluated our framework on ten benchmarks targeting both area and latency optimizations; for each configuration, we ran each benchmark 10 times with each model. 
We used the recommended parameters from each LLM provider: Claude Sonnet 3.5 with \texttt{temperature}=0.2 and \texttt{top\_p}=0.2, and ChatGPT-4o with \texttt{temperature}=0.25 and \texttt{top\_p}=0.2.

\subsection{Benchmark Characterization}
Previous work for Verilog generation is usually evaluated on RTLLM~\cite{rtllm} and/or VerilogEval~\cite{verilogeval}, which are composed of very simple, grad-school class exercise levels at most. High-level synthesis code generation studies like~\cite{eval_llm_hls,automatedhls,amaranth} also focus on small, exercise-like problems.
For our evaluation, we picked 10 benchmarks, of which 9 consist of real-world case applications. We selected three ciphers: AES, DES, and Present. One hashing function, sha256. Five tests for randomness evaluation: monobit, monobit block, cumulative sums, runs, and overlapping input pattern. We selected the quicksort algorithm to showcase an example of a recursive application.
The ciphers and hash functions allow us to test our approach on dataflow-intensive designs with multiple hierarchy levels. The randomness tests allow us to stress our framework with applications that require to be refactored with a streaming interface. 
Both scenarios have in common that the algorithms come from standards that publish reference implementations, that might be wanted to be accelerated in hardware using HLS but are not compatible out of the box.
\autoref{tab:benchmarks_auto} reports the benchmarks with their original sources and summarizes their characteristics. Some benchmarks needed minor modifications to work with our framework; all input code for our evaluation, together with the raw results, is available at our repo: 
\href{https://github.com/Lucaz97/c2hlsc}{C2HLSC}.

\begin{table}[ht]
\caption{Benchmarks characterization. Min. and Max. counts are reported by single functions.}\vspace{-8pt}
\label{tab:benchmarks_auto}
\begin{tabular}{@{}llrrrrrrrr@{}}
\toprule
                &              &            &         & \multicolumn{3}{c}{\# Lines} & \multicolumn{3}{c}{\# Operations} \\ \cmidrule(l){5-10} 
Benchmark       & Feature      & \# Function & \# Calls & Total    & Min.    & Max.    & Total      & Min.      & Max.     \\ \midrule
AES             & Hierarchical & 6          & 10      & 101      & 5       & 26     & 77         & 2         & 29      \\
DES             & Hierarchical & 4          & 4       & 59       & 6       & 24     & 1263       & 6         & 512     \\
present         & Hierarchical & 6          & 9       & 96       & 10      & 24     & 74         & 4         & 26      \\
SHA256          & Hierarchical & 2          & 1       & 70       & 17      & 53     & 127        & 11        & 116     \\
QuickSort       & Recursion    & 3          & 5       & 33       & 6       & 19     & 13         & 3         & 10      \\
Cumulative Sums & Streaming    & 1          & 0       & 22       & -       & -      & 10         & -         & -       \\
Monobit         & Streaming    & 1          & 0       & 11       & -       & -      & 5          & -         & -       \\
Monobit Block   & Streaming    & 1          & 0       & 24       & -       & -      & 10         & -         & -       \\
Overlapping     & Streaming    & 1          & 0       & 45       & -       & -      & 20         & -         & -       \\
Runs            & Streaming    & 1          & 0       & 19       & -       & -      & 10         & -         & -       \\
 
\bottomrule
\end{tabular}
\end{table}

\subsection{Experimental Results}

\autoref{tab:synth_area_openai} and \autoref{tab:synth_lat_openai} report success rate, the number of compile and HLS runs, together with the area and latency values, for the GPT4o/GPT4o-mini ensemble targeting area and latency optimization, respectively. 
In both cases, the ensemble approach using Open AI 4o and 4o-mini models did not succeed at obtaining HLS-compatible code for DES. It succeeded only once for Monobit Block with area target optimizations and once for Overlapping Input Patterns for latency target optimizations. Looking at \autoref{tab:sonnet_area_synth} and \autoref{tab:synth_lat_sonnet}, which report the same data for Sonnet 3.5 targeting area and latency optimization, respectively, we find that Sonnet 3.5 also did not succeed with DES and struggled with Overlapping Input Patterns (1 success out of 10). The DES implementation that we selected uses a lot of pre-compiler macros to implement state operations which results in a C code that is rich in operations (as reflected in \autoref{tab:benchmarks_auto}). Examining the logs for the DES runs, we found that the most common errors were functional errors. We attribute this to the higher complexity of the single functions after the preprocessor expansions of the macros. In order to parse the input code for the hierarchical preprocessing step, we need to run the c preprocessor. This highlights a limitation in the code style for our framework; limiting the use of macros seems to increase the success rate. With the Overlapping Input Pattern and Monobit Block, the challenge lies in the presence of nested loops (3 for the former and 2 for the latter) that need to be flattened for the streaming interface implementation.
The number of compile runs has a higher average value and a wider range compared to the number of HLS runs, suggesting that the generative model struggles more at producing valid and compilable code but is fairly capable of addressing the synthesis issues. These results highlight the importance of the compilation and functional test loop before running the LLM-generated code with HLS. For the area target optimizations, the GPT4o/GPT4o-mini ensemble adopted a strategy of never applying pragmas as pipelining and unrolling increase area, citing a common response: \textit{"To optimize the `AddRoundKey` function for the area, we should avoid loop unrolling and pipelining, as these optimizations typically increase area usage. Instead, we will keep the loops as they are, which is the most area-efficient approach."}. For this reason, area and latency values for the area optimizations target have very low variability. The changes are only due to code structure. CuSums and Runs present a wide range for latency, this is due to the cases in which the model did not implement a proper streaming interface, we will discuss streaming interface results in more details later. For latency optimization, we can spot a lot more variability in the results as the model comes up with different approaches. 
For Claude Sonnet 3.5, the area and latency results for both optimization targets do not present much variability. This is due to Sonnet often hallucinating the syntax for the pragmas, which results in them being ignored by the synthesis tool. For the Streaming interface benchmarks, the high max latency results are due to the model failing to implement a proper streaming interface. In some of the instances that failed to provide a streaming interface, we notice unexpected synthesis results (i.e. 0 latency in cusums, runs and overlapping).\footnote{After analysis, we found that the HLS tool does not report any errors, although the synthesis result is clearly incorrectly reporting zero latency. A closer look shows that the HLS tool misses data dependencies with global arrays. We discarded these results, which we do not consider a success in Tables 3 to 6.}

\begin{table}[tb]
\caption{Success rate, number of compile and HLS runs, and synthesis results from GPT4o/GPT4o-mini ensemble with area optimizations target.} \vspace{-8pt}
\label{tab:synth_area_openai}
\resizebox{\textwidth}{!}{
\begin{tabular}{@{}lrrrrrrrrrrrrr@{}}
\toprule
 & Succ.&\multicolumn{3}{c}{\# Compile Runs} & \multicolumn{3}{c}{\# HLS Runs} & \multicolumn{3}{c}{Area {[}um2{]}} & \multicolumn{3}{c}{Latency {[}cycles{]}} \\ \cmidrule(l){3-14} 
Benchmark & {[}\%{]}& \multicolumn{1}{l}{Avg.} & \multicolumn{1}{l}{Min.} & \multicolumn{1}{l}{Max.} & \multicolumn{1}{l}{Avg.} & \multicolumn{1}{l}{Min.} & \multicolumn{1}{l}{Max.} & \multicolumn{1}{l}{Avg.} & \multicolumn{1}{l}{Min.} & \multicolumn{1}{l}{Max.} & \multicolumn{1}{l}{Avg.} & \multicolumn{1}{l}{Min.} & \multicolumn{1}{l}{Max.} \\ \midrule
AES & 80 & 36.63 & 29 & 44 & 16.00 & 16 & 16 & 2975.90 & 2975.90 & 2975.90 & 853 & 853 & 853 \\
DES & 0 & - & - & - & - & - & - & - & - & - & - & - & - \\
Present & 30& 39.33 & 37 & 41 & 18.00 & 18 & 18 & 19985.20 & 19985.20 & 19985.20 & 6193 & 6193 & 6193 \\
SHA256 & 100 & 12.00 & 11 & 13 & 7.20 & 7 & 8 & 41794.10 & 41794.10 & 41794.10 & 83 & 83 & 83 \\
CuSums & 100 & 6.10 & 4 & 10 & 2.00 & 2 & 2 & 1732.54 & 1522.70 & 2058.50 & 8002 & 1 & 40001 \\
Monobit & 100 & 4.20 & 4 & 5 & 2.00 & 2 & 2 & 808.50 & 808.50 & 808.50 & 1 & 1 & 1 \\
Block & 10 & 14.00 & 14 & 14 & 4.00 & 4 & 4 & 14897.00 & 14897.00 & 14897.00 & 32 & 32 & 32 \\
Overlap. & 0 & - & - & - & - & - & - & - & - & - & - & - & - \\
Runs & 60 & 13.10 & 11 & 19 & 2.00 & 2 & 2 &632.96 & 200.30 & 1453.40 & 23592 & 1 & 65535 \\
Q.S. & 40 & 10.00 & 7 & 18 & 5.00 & 4 & 8 & 47880.30 & 12089.10 & 81122.90 & 4 & 4 & 4 \\ \bottomrule
\end{tabular}
}
\end{table}

\begin{table}[tb]
\caption{Success rate, number of compile and HLS runs, and synthesis results from GPT4o/GPT4o-mini ensemble with latency optimizations target.}\vspace{-8pt}
\label{tab:synth_lat_openai}
\resizebox{\textwidth}{!}{
\begin{tabular}{@{}lrrrrrrrrrrrrr@{}}
\toprule
 & Succ. &\multicolumn{3}{c}{\# Compile Runs} & \multicolumn{3}{c}{\# HLS Runs} & \multicolumn{3}{c}{Area {[}um2{]}} & \multicolumn{3}{c}{Latency {[}cycles{]}} \\ \cmidrule(l){3-14} 
Benchmark & {[}\%{]}&\multicolumn{1}{l}{Avg.} & \multicolumn{1}{l}{Min.} & \multicolumn{1}{l}{Max.} & \multicolumn{1}{l}{Avg.} & \multicolumn{1}{l}{Min.} & \multicolumn{1}{l}{Max.} & \multicolumn{1}{l}{Avg.} & \multicolumn{1}{l}{Min.} & \multicolumn{1}{l}{Max.} & \multicolumn{1}{l}{Avg.} & \multicolumn{1}{l}{Min.} & \multicolumn{1}{l}{Max.} \\ \midrule
AES & 60 & 34.83 & 29 & 44 & 16.00 & 16 & 16 & 17812.60 & 6660.80 & 20920.00 & 231 & 159 & 571 \\
DES & 0 & - & - & - & - & - & - & - & - & - & - & - & - \\
Present &30 & 38.00 & 37 & 39 & 18.33 & 18 & 19 & 22245.17 & 19262.80 & 24953.40 & 897 & 609 & 1154 \\
SHA256 &100 & 12.70 & 9 & 14 & 6.10 & 6 & 7 & 47539.87 & 37894.00 & 51580.90 & 422 & 67 & 573 \\
CuSums &90 & 5.44 & 4 & 12 & 2.00 & 2 & 2 & 1787.14 & 1522.70 & 2309.70 & 11113 & 1 & 60000 \\
Monobit &100 & 4.30 & 4 & 7 & 2.00 & 2 & 2 & 808.50 & 808.50 & 808.50 & 1 & 1 & 1 \\
Block &0 & - & - & - & - & - & - & - & - & - & - & - & - \\
Overlap. & 0 & - & - & - & - & - &  - &- & - & - & - & - & - \\
Runs &70 & 11.56 & 10 & 12 & 2.00 & 2 & 2 & 557.50 & 200.30 & 1452.70 &18725 & 1 & 65536 \\
Q.S. &60 & 9.33 & 7 & 19 & 4.33 & 4 & 6 & 61850.53 & 11922.60 & 83227.00 & 4 & 4 & 4 \\ \bottomrule
\end{tabular}
}\vspace{-6pt}
\end{table}

\begin{table}[tb]
\caption{Success rate, number of compile and HLS runs, and synthesis results from Sonnet 3.5 ensemble with area optimizations target.} \vspace{-8pt}
\label{tab:sonnet_area_synth}
\resizebox{\textwidth}{!}{
\begin{tabular}{@{}lrrrrrrrrrrrrr@{}}
\toprule
 & Succ. &\multicolumn{3}{c}{\# Compile Runs} & \multicolumn{3}{l}{\# HLS Runs} & \multicolumn{3}{c}{Area {[}um2{]}} & \multicolumn{3}{c}{Latency {[}cycles{]}} \\ \cmidrule(l){3-14} 
Benchmark & {[}\%{]} & \multicolumn{1}{l}{Avg.} & \multicolumn{1}{l}{Min.} & \multicolumn{1}{l}{Max.} & \multicolumn{1}{l}{Avg.} & \multicolumn{1}{l}{Min.} & \multicolumn{1}{l}{Max.} & \multicolumn{1}{l}{Avg.} & \multicolumn{1}{l}{Min.} & \multicolumn{1}{l}{Max.} & \multicolumn{1}{l}{Avg.} & \multicolumn{1}{l}{Min.} & \multicolumn{1}{l}{Max.} \\ \midrule
AES & 10 & 55.00 & 55 & 55 & 25.00 & 25 & 25 & 2965.30 & 2965.30 & 2965.30 & 853 & 853 & 853 \\
DES & 0 & - & - & - & - & - & - & - & - & - & - & - & - \\
Present & 40 & 38.25 & 37 & 39 & 18.00 & 18 & 18 & 19566.43 & 18709.60 & 19984.70 & 6367 & 6193 & 6888 \\
SHA256 & 100 & 14.10 & 12 & 24 & 6.20 & 6 & 8 & 41924.00 & 41794.10 & 42227.10 & 83 & 83 & 83 \\
CuSums & 3& 4.70 & 4 & 6 & 2.00 & 2 & 2 & 2083.70 & 1803.20 & 2269.80 & 2 & 1 & 4 \\
Monobit & 100 & 4.00 & 4 & 4 & 2.00 & 2 & 2 & 813.20 & 813.20 & 813.20 & 1 & 1 & 1 \\
Block & 80 & 13.63 & 11 & 18 & 4.13 & 4 & 5 & 2861.36 & 104.40 & 11325.20 & 33 & 1 & 256 \\
Overlap. & 20 & 29.50 & 24 & 35 & 3.50 & 3 & 4 & 17923.35 & 10473.90 & 25372.80 & 17 & 1 & 32 \\
Runs & 100 & 10.00 & 9 & 11 & 2.00 & 2 & 2 & 1958.70 & 1889.30 & 2236.30 & 1 & 1 & 3 \\
Q.S. & 100 & 9.40 & 5 & 17 & 4.60 & 4 & 7 & 53746.85 & 11813.60 & 83659.80 & 4 & 4 & 4 \\ \bottomrule
\end{tabular}
}\vspace{-6pt}
\end{table}

\begin{table}[tb]
\caption{Success rate, number of compile and HLS runs, and synthesis results from Sonnet 3.5 ensemble with latency optimizations target.}\vspace{-8pt}
\label{tab:synth_lat_sonnet}
\resizebox{\textwidth}{!}{
\begin{tabular}{@{}lrrrrrrrrrrrrr@{}}
\toprule
 & Succ. &\multicolumn{3}{c}{\# Compile Runs} & \multicolumn{3}{c}{\# HLS Runs} & \multicolumn{3}{c}{Area {[}um2{]}} & \multicolumn{3}{c}{Latency {[}cycles{]}} \\ \cmidrule(l){3-14} 
Benchmark & {[}\%{]} & \multicolumn{1}{l}{Avg.} & \multicolumn{1}{l}{Min.} & \multicolumn{1}{l}{Max.} & \multicolumn{1}{l}{Avg.} & \multicolumn{1}{l}{Min.} & \multicolumn{1}{l}{Max.} & \multicolumn{1}{l}{Avg.} & \multicolumn{1}{l}{Min.} & \multicolumn{1}{l}{Max.} & \multicolumn{1}{l}{Avg.} & \multicolumn{1}{l}{Min.} & \multicolumn{1}{l}{Max.} \\ \midrule
AES & 80 & 27.29 & 26 & 29 & 19.43 & 19 & 21 & 5523.77 & 4097.20 & 6094.40 & 601 & 572 & 672 \\
DES & 0 & - & - & - & - & - & - & - & - & - & - & - & - \\
Present & 100 & 37.50 & 36 & 41 & 18.30 & 18 & 20 & 21621.76 & 20420.30 & 29182.10 & 954 & 280 & 1529 \\
SHA256 & 100 & 16.00 & 14 & 20 & 6.00 & 6 & 6 & 41890.32 & 41794.10 & 42227.10 & 83 & 83 & 83 \\
Cusums & 100 & 4.00 & 4 & 4 & 2.00 & 2 & 2 & 205.10 & 205.10 & 205.10 & 20001 & 20001 & 20001 \\
Monobit & 100 & 4.00 & 4 & 4 & 2.00 & 2 & 2 & 455.75 & 98.30 & 813.20 & 65 & 1 & 128 \\
Block & 50 & 14.43 & 10 & 20 & 4.14 & 4 & 5 & 3306.53 & 548.10 & 11325.20 & 1 & 1 & 1 \\
Overlap. & 10 & 22.00 & 22 & 22 & 5.00 & 5 & 5 & 38160.00 & 38160.00 & 38160.00 & 583 & 583 & 583 \\
Runs & 90 & 10.10 & 9 & 12 & 2.00 & 2 & 2 & 627.50 & 200.30 & 2050.60 &43691 & 1 & 65536 \\
Q.S. & 100 & 10.00 & 5 & 15 & 4.80 & 4 & 7 & 58270.55 & 11856.90 & 107750.30 & 3 & 1 & 4 \\ \bottomrule
\end{tabular}
}\vspace{-6pt}
\end{table}

\begin{figure}[htb]
    \centering
    \begin{minipage}[b]{\textwidth}
        \centering
        \includegraphics[width=.9\textwidth]{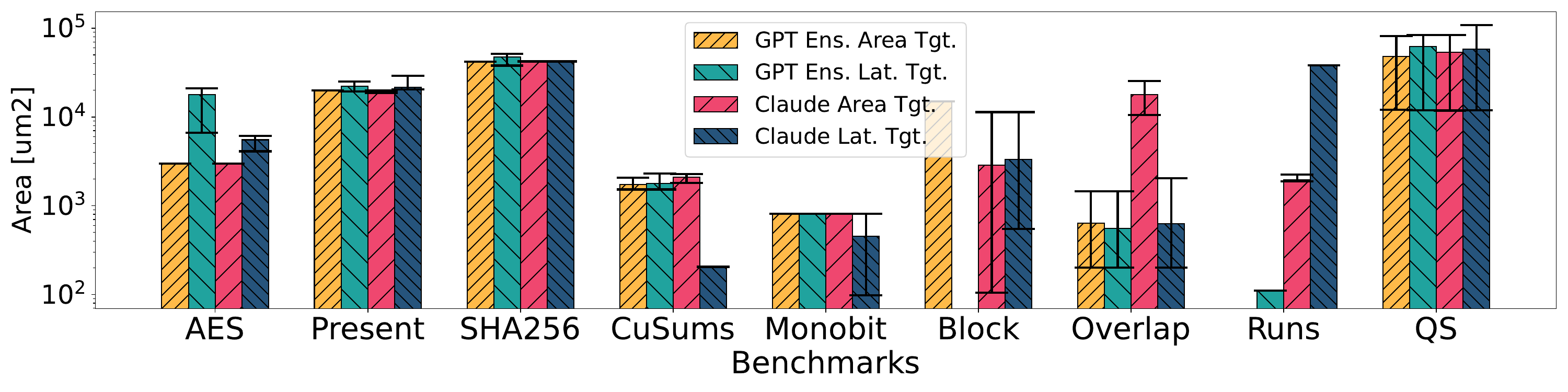}
        \vspace{-10pt}
        \subcaption{Average Area Comparison (log scale)}
        \vspace{5pt}
        \label{fig:subfig3}
    \end{minipage}
    \begin{minipage}[b]{\textwidth}
        \centering
        \includegraphics[width=.9\textwidth]{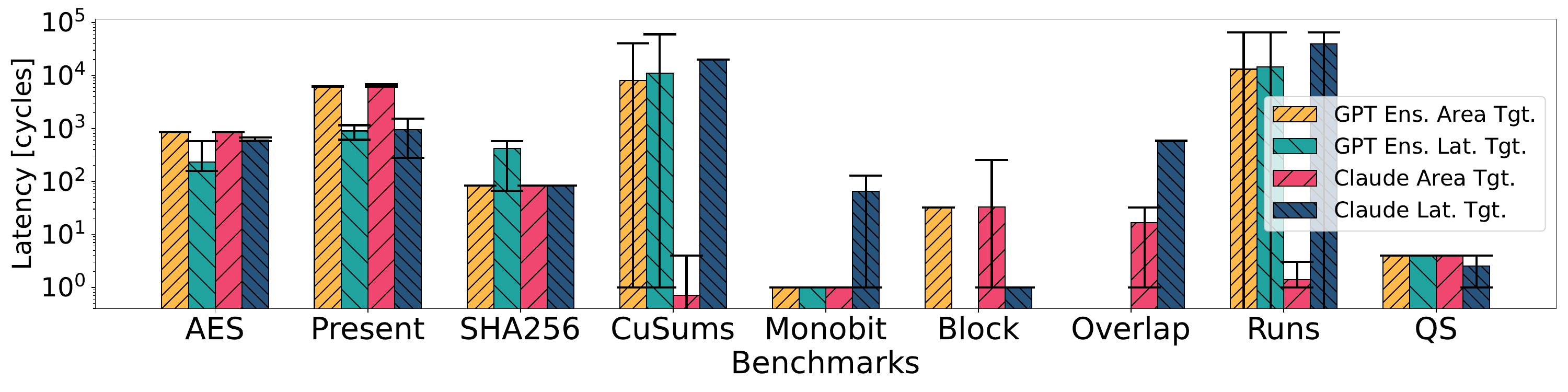}
        
        \subcaption{Average Latency Comparison (log scale)}
        \label{fig:subfig4}
    \end{minipage}
    \caption{Synthesis comparison between OpenAI and Anthropic models. The error bars represent the min/max ranges.}
    \label{fig:synth_comp}
    \vspace{-10pt}
\end{figure}

\autoref{fig:synth_comp} shows a comparison of area and latency between the 4 different setups. The missing bars are due to all tests failing for the specific model/target pair. We can notice the latency is improved to the cost of the area for AES and Present. For the other designs, the average latency is not always improved, but the minimum latency achieved across the 10 runs is equal or improved. Overall, the GPT4o/GPT4o-mini ensemble performed better at optimizing the code for latency than Sonnet 3.5, as the latter hallucinated pragmas syntax.  Overall, the optimization step was not very effective on the single runs, but we noticed improvements on the best across the 10 performed runs. This highlights the limitation of feedback in the single-shot approach for the optimization step. Future works include expanding the optimization step to have a feedback loop with the synthesis results to guide the optimization instead of the single-shot approach used in this work.

\begin{figure}[htb]
    \centering
    \begin{minipage}[b]{.9\textwidth}
        \centering
        \includegraphics[width=\textwidth]{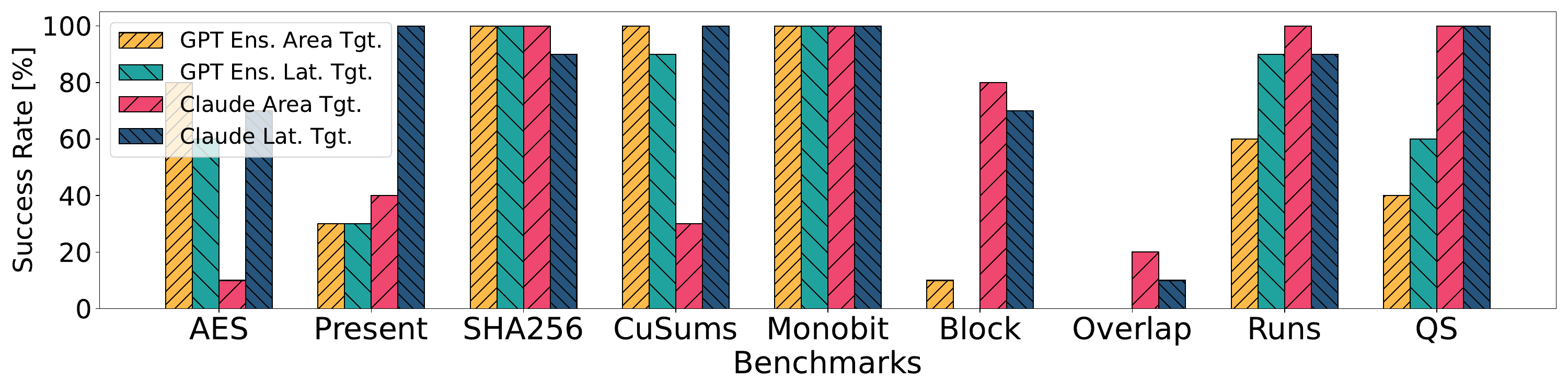}
        \vspace{-18pt}
        \subcaption{Success Rate Comparison)}
        \vspace{5pt}
        \label{fig:subfig3}
    \end{minipage}
    \begin{minipage}[b]{\textwidth}
        \centering
        \includegraphics[width=.9\textwidth]{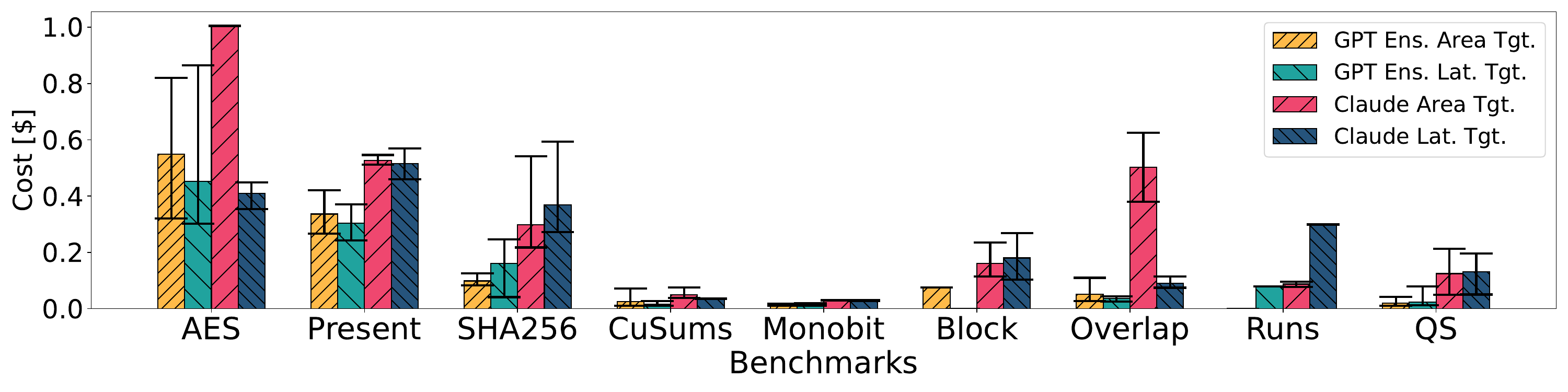}
        \vspace{-9pt}
        \subcaption{Cost Comparison}
        \label{fig:subfig4}
    \end{minipage}
    \caption{Comparisons between OpenAI and Anthropic models success rate and cost. The error bars represent the min/max ranges. The cost was calculated by multiplying the rate of the API provider.}
    \label{fig:cost_sucess}
\end{figure}

\autoref{fig:cost_sucess} compares success rate and cost across models and area/latency targets. The cost was calculated by multiplying the rate of the API provider, distinguishing between input and output tokens when necessary. Overall, we can notice some variability across the area/latency targets, both for success rate and cost. The bars represent the average values, while the error bars represent the range from minimum to maximum.The missing bars are due to all tests failing for the specific model/target pair. Sonnet 3.5 only succeeded in one out of ten runs on AES with the area optimization target but succeeded eight out of ten times with the latency optimization target. Other benchmarks presented variations, but not as big as the AES ones. The data does not show a significant difference in success rate and number of prompts between the area and latency target optimizations, suggesting that the complexity of the task is constant with respect to the target optimization, but our sampling size is too small to draw conclusions on this matter. Overall, Claude Sonnet 3.5 shows a higher success rate and cost compared to the GPT4o/GPT4o-mini ensemble, with AES being the only benchmark on which Sonnet had a lower success rate than the GPT ensemble. More in-depth information about the models' usage is provided in \autoref{app:models}.

As mentioned above, for some of the Streaming interface benchmarks, the LLMs produce a synthesizable code that does not use the streaming interface. \autoref{tab:stream} reports data on streaming interface success for both models. Overall, Sonnet 3.5 is more effective at generating a streaming interface, with the Cumulative Sums test being an outlier. The Sonnet 3.5-generated code structure for Cumulative Sums is very similar across all 20 total runs we performed. This possibly hints at a bias in the training set.

\begin{table}[ht]
\caption{Streaming interface success. The area and Latency target data are combined, as the streaming interface step is independent of the final optimization target.} \vspace{-8pt}
\label{tab:stream}
\begin{tabular}{@{}lrrrr@{}}
\toprule
Benchmark & \multicolumn{1}{c}{\begin{tabular}[c]{@{}c@{}}GPT Ens.\\ Synth Succ. {[}\%{]}\end{tabular}} & \multicolumn{1}{c}{\begin{tabular}[c]{@{}c@{}}Sonnet 3.5\\ Synth Succ. {[}\%{]}\end{tabular}} & \multicolumn{1}{c}{\begin{tabular}[c]{@{}c@{}}GPT Ens.\\ Streaming Succ. {[}\%{]}\end{tabular}} & \multicolumn{1}{c}{\begin{tabular}[c]{@{}c@{}}Sonnet 3.5\\ Streaming Succ. {[}\%{]}\end{tabular}} \\ \midrule
CuSums & 95 & 100 & 70 & 10 \\
Monobit & 100 & 100 & 100 & 80 \\
Block & 10 & 65 & 0 & 60 \\
Overlapping & 10 & 15 & 0 & 15 \\
Runs & 95 & 100 & 55 & 60 \\ \bottomrule
\end{tabular}
\vspace{-10pt}
\end{table}

\begin{figure}[thb]
  \centering
  \begin{tabular}{cc}
  \begin{subfigure}[t]{.44\columnwidth}
    \begin{lstlisting}
GPT4o/GPT4o-mini result for CuSums 
\end{lstlisting}
\vspace{-7pt}
  \end{subfigure} & \hspace{4pt}
   \begin{subfigure}[t]{.44\columnwidth}
    \begin{lstlisting}
Sonnet 3.5 result for CuSums
    \end{lstlisting}
    \vspace{-7pt}

  \end{subfigure}\\
  \begin{subfigure}[t]{.44\columnwidth}
    \begin{lstlisting}[language=c]
void CumulativeSums(int *res_sup, int *res_inf, int epsilon_elem)
{
    static int S = 0;
    static int sup = 0;
    static int inf = 0;
    // Update S based on the current element
    S += (epsilon_elem) ? 1 : -1;
    // Update sup and inf based on the value of S
    if (S > sup)
        sup++;
    if (S < inf) 
        inf--;
    // Set the results after processing all elements
    *res_sup = sup;
    *res_inf = inf;
}
\end{lstlisting}
\vspace{-7pt}
  \end{subfigure}
  & \hspace{4pt}
  \begin{subfigure}[t]{.44\columnwidth}
    \begin{lstlisting}[language=c]
void CumulativeSums(int *res_sup, int *res_inf) {
    static int S = 0;
    static int sup = 0;
    static int inf = 0;
    static int k = 0;
    #pragma hls_pipeline_init_interval 1
    for (int i = 0; i < N; i++) {
        #pragma hls_unroll yes
        int epsilon_element = epsilon[i];
        (epsilon_element) ? (S++) : (S--);
        if (S > sup)
            sup++;
        if (S < inf)
            inf--;
        k++;
    }
    *res_sup = sup;
    *res_inf = inf;
    // Reset static variables for next use
    S = 0;
    sup = 0;
    inf = 0;
    k = 0;
}
...
\end{lstlisting}
\vspace{-7pt}
  \end{subfigure}
  \end{tabular}
  \caption{Results from a Cumulative Sums run for the GPT4o/GPT4o-mini ensemble and Sonnet 3.5 with area target.}
  \label{fig:cusums}
\end{figure}
\autoref{fig:cusums} shows a result for Cumulative Sums for both models with area target.  Epsilon is the current bit from the sequence that is being observed. S represents the sum obtained by summing one or negative one for every observed one or zero, respectively. Sup and inf represent the upper and lower bounds of S for the observed sequence. For this test, the threshold for determining whether the sequence is random or not can be performed directly on the value of sup, inf, and S. For this reason, no further calculation is needed at the end of the sequence. On the left, the GPT ensemble correctly adds the \texttt{epsilon\_elem} to the function parameters, makes S, sup, and inf, local variables static, and removes the for loop operating on epsilon. On the right, Sonnet 3.5 did make the local variables static but failed at adding the \texttt{epsilon\_element} a new parameter and removing the for loop.

Inspecting the generated code, we found that across the same benchmark, solutions from the same model are very similar and follow one to a few different patterns that are different for different models. This may suggest that where a model fails, and another succeeds, it might be more due to differences in the training data than the model design itself.

\begin{table}[!bht]
\centering
\caption{Comparison between fully automated C2HLSC framework, case study (LLM assisted) and manual approach.}
\vspace{-10pt}
\label{tab:manual}
\begin{tabular}{@{}lrrrrrr@{}}
\toprule
\multirow{2}{*}{Design} & \multicolumn{3}{c}{Area Score} &  \multicolumn{3}{c}{Latency} \\ \cmidrule(l){2-7} 
 & \multicolumn{1}{l}{Manual} & \multicolumn{1}{l}{LLM Assisted} & \multicolumn{1}{l}{C2HLSC} & \multicolumn{1}{l}{Manual} & \multicolumn{1}{l}{LLM Assisted} & \multicolumn{1}{l}{C2HLSC}  \\ \midrule
Monobit & 225.3 & 244.0 & 808.5  & 1 & 1 & 1 \\
Block & 826.0 & 702.3 & 1656.6  & 1 & 1 & 1 \\
Cusums & 632.0 & 677.4 & 1522.7  & 1 & 1 & 1 \\
Overlapping & 7172.1 & 9933.4 & 10473.9 & 1 & 1 & 1 \\ 
Q.S. & - & 18115.8 & 11813.6 & - & 160 & 4 \\
AES & 3386.0 & - & 2965.3 & 193 & - & 853\\
SHA256 & 36090.0 & - & 41890.3 & 48 & - &  83 \\
Present & 12056.0 & - & 21621.7 &  37 & - & 954 \\\bottomrule
\end{tabular}
\end{table}

\autoref{tab:manual} shows a comparison with the overlapping designs of the case study and manual implementations coming from a previous grad student's work. We report the best result from our framework runs. Area and latency improvements are achievable with manual effort.  The discrepancy in area for monobit, block, cusums, and overlapping is due to the use of custom data type optimizations in the case study and manual implementations. This shows that to further improve performance, custom data type optimizations would bring great advantages. On the other hand, the manual implementations took 4-6 hours each; in our LLM-assisted case study (presented in \autoref{sec:cs}), it took around 1 hour each. With our fully automated flow, it takes from a few minutes to 15 minutes, during which the user can be working on a different task. This means that even at the current stage, the designer can invest some time into doing final tweaks to optimize the design. As the cost of one of our runs is one dollar or less, and more powerful models are being released at lower prices, the proposed framework proves to be helpful in reducing costs and time for accelerating C code bases in hardware.

\subsection{Comparison with State of the Art}
Looking at GPT4AIGChip~\cite{GPT4AIGChip} we cannot do direct comparisons due to different benchmark selections, although we can point out that GPT4AIGChip needs a considerable amount of human involvement, whereas our flow is fully automated. Liao et al.~\cite{llm_hls1} investigated using LLMs for C to Verilog transpilation. They used smaller benchmarks than the ones we used in this work. HLSPilot~\cite{llm_hls2} takes as input synthesizable C code and focuses on the optimization, as opposed to the main contribution of our work which consists in refactoring C code such that can be synthesized.
Xu te al.~\cite{automatedhls} is the closest framework in the literature. The only common benchmark is AES. In AES we performed generally better 60\% vs. 80\% (only our Sonnet 3.5 with area target had a lower success rate). We focus not only on the repairs, but also on implementing streaming interfaces and applying pragmas. Overall, our benchmarks are the most complex designs generated and optimized completely automatically, highlighting the benefits of the hierarchical approach and decoupling of the code refactoring phase and the optimization phase.

\section{Conclusions} \label{sec:conc}
With our case study, we demonstrated the potential of LLMs in bridging the software-to-hardware design gap by leveraging HLS. With the hindsight gathered in the case study, we implemented a fully automated framework that is able to take complex C code and rewrite it in a way that is compatible with HLS tools using LLMs. To achieve this, we break down the design hierarchy and approach the rewrite process in a bottom-up approach. Once the code is synthesizable, we task the LLM to add pragmas to optimize the hardware implementation obtained with the HLS tool. 
Our results show that the approach is very effective at generating HLS synthesizable code. We obtain a good success rate on designs that are the order of magnitude more complex than what has been achieved when generating Verilog directly from the LLMs. This validates our hypothesis that working at a higher level of abstraction would allow us to better use LLMs for hardware generation. The LLMs struggled in the optimization step. On one side, more in-context learning or retrieval-augmented generation techniques could be employed to improve the results. On the other side, pragma exploration is a much more constrained problem than code rewriting, and might be more effective in tackling this problem without LLMs. Our work is open-source and available at \href{https://github.com/Lucaz97/c2hlsc}{C2HLSC}. Future work includes expanding to more complex code bases, handling C++ object-oriented constructs, and improving the optimization pragma insertion phase by using synthesis feedback and RAG mechanisms for pragma syntax.

\bibliographystyle{ACM-Reference-Format}
\bibliography{main}

\newpage
\appendix

\section{Prompts and In-Context Learning (ICL) Examples}

\subsection{Code Refactoring ICL} \label{sec:refactoring_icl}
In the code refactoring step (\autoref{sec:refactor}), we provide examples for in-context learning to obtain a streaming interface and fix the following errors:
\begin{itemize}
    \item \textbf{Streaming interface}:  
\begin{lstlisting}
Rewrite the {top_function} function to be compatible for HLS. The first task is to rewrite it
such that it will get inferred as a streaming function, to do so, I need to get rid of the 
global array and have the function take a parameter to accept one element at each function call
The following is an example on how this can be done: 
```
#define N 20
#define TAPS 11
int x[N];
void fir(*y) {
    int  c[TAPS] = { 53, 0, -91, 0, 313, 500, 313, 0, -91, 0, 53};
    static int shift_reg[TAPS];
    int acc;
    int i, j;
    acc = 0;
    for (j = 0; j < N; j++) {
        for (i = TAPS - 1; i >= 0; i--) {
            if (i == 0) {
                acc += x[j] * c[0];
                shift_reg[0] = x[j];
            } else {
                shift_reg[i] = shift_reg[i - 1];
                acc += shift_reg[i] * c[i];
            }
        }
    }
    *y = acc;
}
// Streaming function
#define TAPS 11
void fir(int *y, int x) { // takes one element of x and produces one element of y at each function call 
    int  c[TAPS] = { 53, 0, -91, 0, 313, 500, 313, 0, -91, 0, 53};
    static int shift_reg[TAPS]; // this needs to be static to be preserved across function calls
    static int acc;
    static int j = 0;
    int i;
    acc = 0;
    for (i = TAPS - 1; i >= 0; i--) {
        if (i == 0) {
            acc += x * c[0];
            shift_reg[0] = x;
        } else {
            shift_reg[i] = shift_reg[i - 1];
            acc += shift_reg[i] * c[i];
        }
    }
    if (j==N) {
        *y = acc;
        j = 0;
    } else {
        j++;
    }
}
```
If there is more than one loop one will need multiple if statements to differentiate the outer loops actions.
The final function must not contain loops.
        
        
\end{lstlisting}
    \item \textbf{Recursion}:

        \begin{lstlisting}

Here are two examples on how simple cases and more complex cases of recursion can be rewritten to avoid recursion:
Tail recursive function
```
algorithm SolveTailRecursive(problem, accumulator):
    // INPUT
    //    problem = an instance of the problem to solve
    //    accumulator = the variable that holds partial solutions
    // OUTPUT
    //    solution = the complete solution to the problem or an indicator that no solution exists
    if BaseCase(problem):
        accumulator <- apply the base-case update
        return accumulator
    else:
        // body
        accumulator <- update the accumulator
        subproblem <- reduce problem to a smaller sub-problem
        return SolveTailRecursive(subproblem, accumulator)
```
Iterative version:
```
algorithm SolveTailIterative(problem):
    // INPUT
    //    problem = an instance of the problem to solve
    // OUTPUT
    //    solution = the complete solution to the problem (or an indicator that no solution exists)
    accumulator <- initialize the accumulator
    while not BaseCase(problem):
        accumulator <- update the accumulator
        subproblem <- reduce problem to a smaller sub-problem
        problem <- subproblem
    accumulator <- apply the base-case update
    return accumulator
```
General recursive case:
```
algorithm SolveRecursive(problem):
    // INPUT
    //    problem = an instance of problem to solve
    // OUTPUT
    //    The solution to problem if one exists, or failure - notification of its inexistence, otherwise
    if BaseCase(problem):
        return the base-case solution to problem
    else:
        i <- 0
        while there is a recursive call to make:
            i <- i + 1
            Execute NRCB_i, the non-recursive code block number i
            subproblem_i <- extract the i-th sub-problem from problem
            subsolution_i <- SolveRecursive(subproblem_i)
    
        // let m be the number of sub-solutions (and sub-problems)
        solution <- combine subsolution_1, ..., subsolution_m
        if solution is valid:
            return solution
        else:
            return failure
```
General Iterative version:
```
algorithm SolveIter(problem):
    // INPUT
    //    problem = an instance of the problem to solve
    // OUTPUT
    //    The solution to problem if one exists, or failure - notification of its inexistence, otherwise
    start <- CreateFrame(problem)
    start.parent <- NONE
    stack <- create a stack with start as its only element
    while stack is not empty:
        frame <- pop the top element from stack
        if frame has an unvisited out-going edge:
            edge <- GetNextEdge(frame)
            Execute edge.NRCB
            Push frame onto stack
            Push edge.child onto stack
        else:
            solution <- GetReturnValue(frame)
            if frame.parent != NONE:
                Pass the return value of frame to frame.parent
    return GetReturnValue(start)
```
\end{lstlisting}
    \item \textbf{Pointer in the interface}:
    \begin{lstlisting}
You can get rid of pointers in the interface using the array notation like
void foo(int a[SIZE]);
you will need to substitute SIZE with the size of the array. 
In the usage of the parameter a you can use the array notation as well, 
like a[i] instead of *a[i].
    \end{lstlisting}
    \item \textbf{Floating point use}:
    \begin{lstlisting}
You can replace floating point types with ac_fixed or ac_float types.
ac_fixed:
ac_fixed<W, I, false> unsigned fixed-point type with W total bits and I integer bits.
ac_fixed<W, I, true> signed fixed-point type with W total bits and I integer bits.
ac_float:
ac_float<W,I,E,Q>
where the first two parameters W and I define the mantissa as an ac_fixed<W,I,true>, 
the E defines the exponent as an ac_int<E,true> and Q defines the rounding mode
you do not need to include any lib I will include the following:
#include "../include/ac_float.h"
#include "../include/ac_fixed.h
    \end{lstlisting}
    \item \textbf{Redefinition of function}:
    \begin{lstlisting}
To solve this problem you can get rid of the function in the error
as I have already defined it in my code.
    \end{lstlisting}
\end{itemize}

\section{LLM prompt data} \label{app:models}

Here we report the LLM usage data for the fully automated framework experiments. For the GPT ensemble, we can see that the usage of the 4o-mini model is lower than that of the 4o model. This is because we switch from 4o-mini to 4o after three consecutive errors. This is an attempt to save cost as described in \autoref{sec:exp_eval}. In many benchmarks the 4o-mini model fails quickly without recovery and is switched out, leading to low number of prompts and tokens in \autoref{tab:prompts_area_openai}  and \autoref{tab:gpt_lat_target}. Overall, we can notice how the number of prompts and used tokens correlates with the complexity of the benchmark, as benchmarks with higher failure rates have higher LLM usage.
\begin{table}[hb]
\caption{LLM usage data from GPT4o/GPT4o-mini ensemble with area optimizations target.} \vspace{-8pt}
\label{tab:prompts_area_openai}
\resizebox{\textwidth}{!}{
\begin{tabular}{@{}lrrrrrrrrrrrrrrrrrrr@{}}
\toprule
 &  \multicolumn{6}{c}{\# LLM Promts} & \multicolumn{6}{c}{\# Input Tokens} & \multicolumn{6}{c}{\# Output Tokens} \\ \cmidrule(l){2-19} 
 & \multicolumn{3}{c|}{4o} & \multicolumn{3}{c|}{4o-mini} & \multicolumn{3}{c|}{4o} & \multicolumn{3}{c|}{4o-mini} & \multicolumn{3}{c|}{4o} & \multicolumn{3}{c}{4o-mini} \\ \cmidrule(l){2-19} 
Benchmark &  \multicolumn{1}{l}{Avg.} & \multicolumn{1}{l}{Min} & \multicolumn{1}{l|}{Max.} & \multicolumn{1}{l}{Avg.} & \multicolumn{1}{l}{Min.} & \multicolumn{1}{l|}{Max.} & \multicolumn{1}{l}{Avg.} & \multicolumn{1}{l}{Min.} & \multicolumn{1}{l|}{Max.} & \multicolumn{1}{l}{Avg.} & \multicolumn{1}{l}{Max.} & \multicolumn{1}{l|}{Max.} & \multicolumn{1}{l}{Avg.} & \multicolumn{1}{l}{Min.} & \multicolumn{1}{l|}{Max.} & \multicolumn{1}{l}{Avg.} & \multicolumn{1}{l}{Max.} & \multicolumn{1}{l}{Max.} \\ \midrule
AES   & 18.25 & 13 & \multicolumn{1}{r|}{24} & 6.75 & 5 & \multicolumn{1}{r|}{9} & 117481 & 61078 & \multicolumn{1}{r|}{192183} & 25500 & 11863 & \multicolumn{1}{r|}{40455} & 24656 & 16408 & \multicolumn{1}{r|}{32561} & 8106 & 3326 & 13314 \\
DES  & - & - & \multicolumn{1}{r|}{-} & - & - & \multicolumn{1}{r|}{-} & - & - & \multicolumn{1}{r|}{-} & - & - & \multicolumn{1}{r|}{-} & - & - & \multicolumn{1}{r|}{-} & - & - & - \\
Present  & 18.33 & 16 & \multicolumn{1}{r|}{20} & 9.00 & 9 & \multicolumn{1}{r|}{9} & 64909 & 48302 & \multicolumn{1}{r|}{88661} & 19465 & 19336 & \multicolumn{1}{r|}{19536} & 16659 & 13825 & \multicolumn{1}{r|}{19108} & 7467 & 7383 & 7516 \\
SHA256  & 5.80 & 5 & \multicolumn{1}{r|}{7} & 2.20 & 2 & \multicolumn{1}{r|}{3} & 15043 & 12097 & \multicolumn{1}{r|}{20615} & 4709 & 4113 & \multicolumn{1}{r|}{7092} & 5889 & 5075 & \multicolumn{1}{r|}{7066} & 2262 & 2057 & 3053 \\
CuSums  & 3.60 & 2 & \multicolumn{1}{r|}{8} & 1.50 & 1 & \multicolumn{1}{r|}{3} & 4582 & 1403 & \multicolumn{1}{r|}{16309} & 1733 & 957 & \multicolumn{1}{r|}{4180} & 1334 & 666 & \multicolumn{1}{r|}{2989} & 550 & 340 & 1090 \\
Monobit  & 2.20 & 2 & \multicolumn{1}{r|}{3} & 1.00 & 1 & \multicolumn{1}{r|}{1} & 1593 & 1319 & \multicolumn{1}{r|}{2604} & 870 & 870 & \multicolumn{1}{r|}{870} & 704 & 606 & \multicolumn{1}{r|}{1001} & 287 & 274 & 304 \\
Block  & 8.00 & 8 & \multicolumn{1}{r|}{8} & 4.00 & 4 & \multicolumn{1}{r|}{4} & 12675 & 12675 & \multicolumn{1}{r|}{12675} & 4951 & 4951 & \multicolumn{1}{r|}{4951} & 4199 & 4199 & \multicolumn{1}{r|}{4199} & 2073 & 2073 & 2073 \\
Overlap.  & - & - & \multicolumn{1}{r|}{-} & - & - & \multicolumn{1}{r|}{-} & - & - & \multicolumn{1}{r|}{-} & - & - & \multicolumn{1}{r|}{-} & - & - & \multicolumn{1}{r|}{-} & - & - & - \\
Runs  & 5.90 & 4 & \multicolumn{1}{r|}{11} & 4.00 & 4 & \multicolumn{1}{r|}{4} & 10651 & 4844 & \multicolumn{1}{r|}{26781} & 6853 & 6593 & \multicolumn{1}{r|}{7067} & 2209 & 1374 & \multicolumn{1}{r|}{4048} & 1751 & 1544 & 2027 \\
Q.S.  & 1.75 & 1 & \multicolumn{1}{r|}{4} & 5.25 & 4 & \multicolumn{1}{r|}{8} & 1740 & 833 & \multicolumn{1}{r|}{4378} & 7179 & 5218 & \multicolumn{1}{r|}{10805} & 1174 & 656 & \multicolumn{1}{r|}{2650} & 3303 & 2530 & 4899 \\ \bottomrule
\end{tabular}
}
\end{table}

\begin{table}[hb]
\caption{LLM usage data from GPT4o/GPT4o-mini ensemble with latency optimizations target.}\vspace{-8pt}
\label{tab:gpt_lat_target}
\resizebox{\textwidth}{!}{
\begin{tabular}{@{}lrrrrrrrrrrrrrrrrrrr@{}}
\toprule
  & \multicolumn{6}{c}{\# LLM Promts} & \multicolumn{6}{c}{\# Input Tokens} & \multicolumn{6}{c}{\# Output Tokens} \\ \cmidrule(l){2-19} 
  & \multicolumn{3}{c|}{4o} & \multicolumn{3}{c|}{4o-mini} & \multicolumn{3}{c|}{4o} & \multicolumn{3}{c|}{4o-mini} & \multicolumn{3}{c|}{4o} & \multicolumn{3}{c}{4o-mini} \\ \cmidrule(l){2-19} 
Benchmark & \multicolumn{1}{l}{Avg.} & \multicolumn{1}{l}{Min} & \multicolumn{1}{l|}{Max.} & \multicolumn{1}{l}{Avg.} & \multicolumn{1}{l}{Min.} & \multicolumn{1}{l|}{Max.} & \multicolumn{1}{l}{Avg.} & \multicolumn{1}{l}{Min.} & \multicolumn{1}{l|}{Max.} & \multicolumn{1}{l}{Avg.} & \multicolumn{1}{l}{Max.} & \multicolumn{1}{l|}{Max.} & \multicolumn{1}{l}{Avg.} & \multicolumn{1}{l}{Min.} & \multicolumn{1}{l|}{Max.} & \multicolumn{1}{l}{Avg.} & \multicolumn{1}{l}{Max.} & \multicolumn{1}{l}{Max.} \\ \midrule
AES  & 14.50 & 11 & \multicolumn{1}{r|}{21} & 8.33 & 6 & \multicolumn{1}{r|}{11} & 93833 & 56415 & \multicolumn{1}{r|}{209001} & 36089 & 16354 & \multicolumn{1}{r|}{62375} & 20542 & 15631 & \multicolumn{1}{r|}{31980} & 11317 & 4352 & 21587 \\
DES  & - & - & \multicolumn{1}{r|}{-} & - & - & \multicolumn{1}{r|}{-} & - & - & \multicolumn{1}{r|}{-} & - & - & \multicolumn{1}{r|}{-} & - & - & \multicolumn{1}{r|}{-} & - & - & - \\
Present  & 17.00 & 16 & \multicolumn{1}{r|}{18} & 9.00 & 9 & \multicolumn{1}{r|}{9} & 58337 & 39908 & \multicolumn{1}{r|}{80857} & 20076 & 19533 & \multicolumn{1}{r|}{21153} & 15080 & 13607 & \multicolumn{1}{r|}{15923} & 7959 & 7380 & 9040 \\
SHA256  & 6.20 & 3 & \multicolumn{1}{r|}{8} & 2.50 & 2 & \multicolumn{1}{r|}{5} & 29474 & 6380 & \multicolumn{1}{r|}{47648} & 6064 & 4109 & \multicolumn{1}{r|}{17368} & 8387 & 2318 & \multicolumn{1}{r|}{12021} & 2663 & 1987 & 6244 \\
CuSums  & 2.22 & 2 & \multicolumn{1}{r|}{3} & 1.56 & 1 & \multicolumn{1}{r|}{4} & 2126 & 1439 & \multicolumn{1}{r|}{4708} & 1983 & 957 & \multicolumn{1}{r|}{6843} & 961 & 719 & \multicolumn{1}{r|}{1345} & 669 & 320 & 1998 \\
Monobit  & 2.10 & 2 & \multicolumn{1}{r|}{3} & 1.20 & 1 & \multicolumn{1}{r|}{3} & 1517 & 1358 & \multicolumn{1}{r|}{2708} & 1168 & 870 & \multicolumn{1}{r|}{3854} & 730 & 658 & \multicolumn{1}{r|}{1069} & 364 & 274 & 1053 \\
Block  & - & - & \multicolumn{1}{r|}{-} & - & - & \multicolumn{1}{r|}{-} & - & - & \multicolumn{1}{r|}{-} & - & - & \multicolumn{1}{r|}{-} & - & - & \multicolumn{1}{r|}{-} & - & - & - \\
Overlap. & 6.00 & 6 & \multicolumn{1}{r|}{6} & 4.00 & 4 & \multicolumn{1}{r|}{4} & 15675 & 15675 & \multicolumn{1}{r|}{15675} & 8782 & 8782 & \multicolumn{1}{r|}{8782} & 3757 & 3757 & \multicolumn{1}{r|}{3757} & 2660 & 2660 & 2660 \\
Runs  & 4.56 & 4 & \multicolumn{1}{r|}{5} & 3.89 & 3 & \multicolumn{1}{r|}{4} & 6619 & 4444 & \multicolumn{1}{r|}{8369} & 6358 & 3926 & \multicolumn{1}{r|}{7015} & 1779 & 1247 & \multicolumn{1}{r|}{2106} & 1543 & 939 & 1870 \\
Q.S.  & 2.00 & 1 & \multicolumn{1}{r|}{7} & 4.83 & 4 & \multicolumn{1}{r|}{8} & 3049 & 889 & \multicolumn{1}{r|}{13813} & 6660 & 5323 & \multicolumn{1}{r|}{10654} & 1342 & 770 & \multicolumn{1}{r|}{3976} & 3095 & 2639 & 4257 \\ \bottomrule
\end{tabular}
}
\end{table}

\begin{table}[hb]
\caption{LLM usage data from Claude Sonnet 3.5 with area optimizations target.}
\vspace{-8pt}
\label{tab:area_sonnet}
\begin{tabular}{@{}lrrr|rrrrrr@{}}
\toprule
 &  \multicolumn{3}{c|}{\# Sonnet Prompts} & \multicolumn{3}{c}{\# Input Tokens} & \multicolumn{3}{c}{\# Output Tokens} \\ \cmidrule(l){2-10} 
Benchmark  & \multicolumn{1}{l}{Avg.} & \multicolumn{1}{l}{Min.} & \multicolumn{1}{l|}{Max.} & \multicolumn{1}{l}{Avg.} & \multicolumn{1}{l}{Min.} & \multicolumn{1}{l|}{Max.} & \multicolumn{1}{l}{Avg.} & \multicolumn{1}{l}{Min.} & \multicolumn{1}{l}{Max.} \\ \midrule
AES  & 35.00 & 35 & 35 & 117625 & 117625 & \multicolumn{1}{r|}{117625} & 43437 & 43437 & 43437 \\
DES  & - & - & - & - & - & \multicolumn{1}{r|}{-} & - & - & - \\
Present  & 25.25 & 24 & 26 & 53946 & 51704 & \multicolumn{1}{r|}{56049} & 24256 & 23817 & 25190 \\
SHA256  & 9.90 & 8 & 18 & 37691 & 26184 & \multicolumn{1}{r|}{76115} & 12341 & 9246 & 20894 \\
CuSums  & 3.70 & 3 & 5 & 4697 & 3247 & \multicolumn{1}{r|}{8139} & 2357 & 1942 & 3381 \\
Monobit  & 3.00 & 3 & 3 & 2670 & 2640 & \multicolumn{1}{r|}{2694} & 1453 & 1420 & 1538 \\
Block  & 11.38 & 9 & 15 & 19144 & 11873 & \multicolumn{1}{r|}{32406} & 6853 & 5226 & 9199 \\
Overlap.  & 19.50 & 16 & 23 & 76457 & 46598 & \multicolumn{1}{r|}{106315} & 18161 & 15971 & 20350 \\
Runs  & 6.50 & 6 & 7 & 10495 & 8983 & \multicolumn{1}{r|}{11872} & 3732 & 3352 & 4051 \\
Q.S.  & 6.80 & 3 & 12 & 13928 & 3233 & \multicolumn{1}{r|}{23536} & 5517 & 2605 & 9511 \\ \bottomrule
\end{tabular}
\end{table}

\begin{table}[hb]
\caption{LLM usage data from Claude Sonnet 3.5 with latency optimizations target.}
\vspace{-8pt}
\label{tab:runs_claude_lat}
\begin{tabular}{@{}lrrrrrrrrrr@{}}
\toprule
 &  \multicolumn{3}{c}{\# Sonnet Prompts} & \multicolumn{3}{c}{\# Input Tokens} & \multicolumn{3}{c}{\# Output Tokens} \\ \cmidrule(l){2-10} 
Benchmark  & \multicolumn{1}{l}{Avg.} & \multicolumn{1}{l}{Min.} & \multicolumn{1}{l|}{Max.} & \multicolumn{1}{l}{Avg.} & \multicolumn{1}{l}{Min.} & \multicolumn{1}{l|}{Max.} & \multicolumn{1}{l}{Avg.} & \multicolumn{1}{l}{Min.} & \multicolumn{1}{l}{Max.} \\ \midrule
AES & 18.62 & 16 & \multicolumn{1}{r|}{22} & 61428 & 51741 & \multicolumn{1}{r|}{73826} & 19642 & 15623 & 24000 \\
DES   & - & - & \multicolumn{1}{r|}{0} & 0 & - & \multicolumn{1}{r|}{-} & - & - & - \\
Present &  24.10 & 21 & \multicolumn{1}{r|}{26} & 52430 & 41407 & \multicolumn{1}{r|}{67255} & 22589 & 17761 & 26440 \\
SHA256 &  11.10 & 7 & \multicolumn{1}{r|}{19} & 45825 & 19849 & \multicolumn{1}{r|}{97849} & 13925 & 8237 & 22564 \\
CuSums &  3.40 & 3 & \multicolumn{1}{r|}{4} & 4023 & 3176 & \multicolumn{1}{r|}{5241} & 2084 & 1768 & 2498 \\
Monobit &  3.00 & 3 & \multicolumn{1}{r|}{3} & 2699 & 2682 & \multicolumn{1}{r|}{2743} & 1364 & 1341 & 1418 \\
Block & 15.80 & 11 & \multicolumn{1}{r|}{25} & 27537 & 17770 & \multicolumn{1}{r|}{45878} & 9814 & 6481 & 15608 \\
Overlap. & 17.00 & 17 & \multicolumn{1}{r|}{17} & 58951 & 58951 & \multicolumn{1}{r|}{58951} & 15268 & 15268 & 15268 \\
Runs &  6.50 & 6 & \multicolumn{1}{r|}{7} & 10608 & 9061 & \multicolumn{1}{r|}{12255} & 3637 & 3295 & 3995 \\
Q.S. &  6.50 & 3 & \multicolumn{1}{r|}{11} & 13431 & 3285 & \multicolumn{1}{r|}{24492} & 5387 & 2710 & 8747 \\ \bottomrule
\end{tabular}
\end{table}

\end{document}